\documentclass[aip,reprint,nofootinbib]{revtex4-1}
\usepackage{amsmath}
\usepackage{graphics}

\newcommand{\dd}{\mathrm{d}}
\newcommand{\mean}[1]{\left\langle #1 \right\rangle}
\begin{document}


\title[Non-ideal mixing of lipids]{Non-ideal mixing of lipids: a molecular dynamics perspective}



\author{L.Berezovska}
\affiliation{Centre de Biologie Structurale - INSERM -  U 1054 29 rue de Navacelles, 34090 Montpellier, France}
\author{R.Kociurzynski}
\affiliation{Faculty of Biology and University Hospital, Albert Ludwigs University Freiburg, Freiburg, Germany}
\author{F.Thalmann}
\email[]{fabrice.thalmann@ics-cnrs.unistra.fr}
\affiliation{Institut Charles Sadron - CNRS - UPR22
23 rue du Loess, BP 84047, 67034 Strasbourg Cedex 2, France}


\date{\today}

\begin{abstract}
Lipid membranes have complex compositions and modeling the thermodynamic properties of multi-component lipid systems remains a remote goal. In this work we attempt to describe the thermodynamics of binary lipid mixtures by  mapping coarse-grained molecular dynamics systems to two-dimensional simple fluid mixtures. By computing and analyzing the density fluctuations of this model lipid bilayer we determine the numerical value of the quadratic coupling term appearing in a model of regular solutions for the DPPC-DLiPC pair of lipids at three different compositions. Our methodology is general and discussed in detail. 
\end{abstract}


\maketitle 


\section{Introduction}
\subsection{Position of the problem}
Biological membranes lie at the heart of the organization of living cells. Their thin planar geometry arises from the bilayer arrangement of amphiphilic lipid molecules around which the other functional components of the membranes are laid, and that we simply refer to in this work as \textit{lipids} or \textit{lipid molecules}. Membranes containing only self-assembled lipids in water are then called \textit{lipid bilayers}. 

Common membrane forming lipids are typically composed of one hydrophilic headgroup and two hydrophobic alkyl or acyl chains connected by a backbone. There are many different lipid molecules varying in shape, size, charge, chemical headgroup nature  conferring to the membranes to which they belong a wide range of different physical and chemical properties. Lipid bilayers of realistic biological content are therefore  fluids with complex compositions. 

\begin{figure}
\resizebox{0.44\textwidth}{!}{\includegraphics{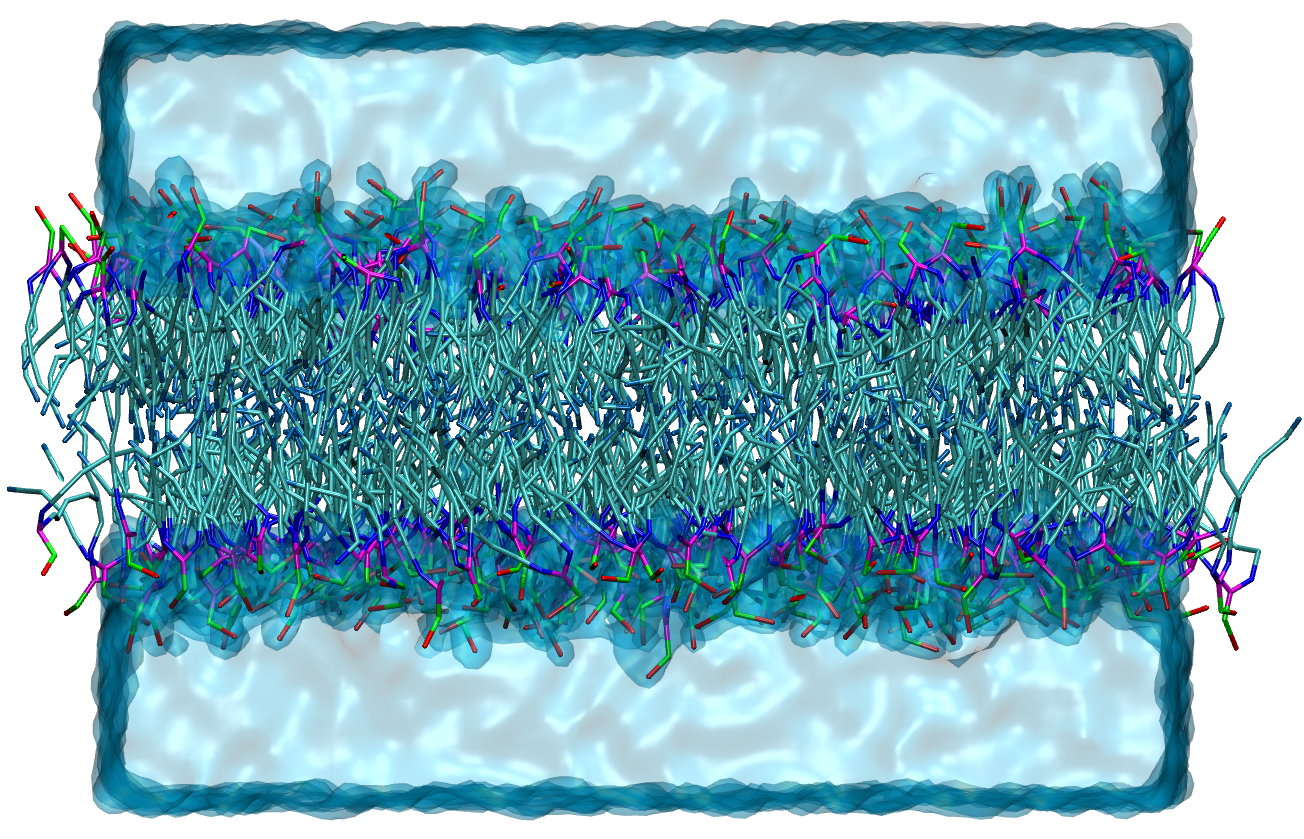}}%
\caption{\label{fig:SnapshotDPPC} Snapshot of a pure DPPC bilayer in fluid state containing 256 lipid molecules per leaflet. The water solvent molecules are not explicitly shown and correspond to the blue shaded regions. }%
\end{figure}

Binary lipid systems only form bilayers of homogeneous composition if the pair of compounds are sufficiently similar in size, shape and chemical composition. For instance, lipids with large chain length discrepancy are expected to segregate into stable coexisting phase domains \cite{1970_Phillips_Chapman,1973_Shimshick_McConnell,1976_Mabrey_Sturtevant, 1988_Ipsen_Mouritsen}. Mixing of lipids in bilayers have been extensively studied and many experimental phase diagrams are now available (see \textit{e.g.}~\cite{Marsh_HandbookLipidBilayers2} and reference therein). A class of ternary lipid mixtures has in particular received much attention~\cite{2009_Marsh} in connection with the problem of lateral segregation of lipids and protein complexes in the plasma membranes of eukaryotic cells (\textit{lipid rafts}). The thermodynamics of mixing of lipids and its relevance to cell membrane biophysics have been reviewed and discussed by many authors~\cite{Heimburg_BiophysicsMembrane, Mouritsen_Bagatolli_MatterOfFat}. 

It is natural to model the properties of bilayer forming lipid mixtures by analogy with the thermodynamics of solutions~\cite{Kirkwood_Oppenheim_ChemicalThermodynamics, BenNaim_WaterSolutions,Atkins_PhysicalChemistry}. The theory of solutions that  describes three dimensional multicomponent fluids  can be transposed to two dimensional systems without difficulty. Then appear two potential obstacles. First, as its name indicates, a bilayer comprises two apposed copies of a thin, fluid leaflet. Each leaflet has a thickness of approx. 1.5-2.5~nm, equal to the normal extension of typical lipids and commensurate to the lateral intermolecular separation ($\sim 0.8~\mathrm{nm}$). This situation can be taken into account by means of a replication of the 2d fluid provided the bilayer has symmetric leaflet composition. Second, free bilayers are not planar but corrugated as a result of their bending elasticity and thermal fluctuations (Helfrich undulations). Though lipid bilayers are stiff enough to be assimilated to planar objects on the smallest length scales, the effect of undulations may become severe on larger length scales. The extent to which the planar fluid approximation describes correctly the properties of lipid bilayers is discussed in the present work. We therefore assume in what follows that a 2d adaptation of the theory of solutions is a valid starting point for discussing the thermodynamics of multicomponent lipid bilayers. 

A key concept of the thermodynamics of solutions is the free-energy of mixing $G^{\mathrm{mix}}$ defined as the free-energy difference between a system with mixed components and a system of identical composition made of juxtaposed, pure, separated molecules in the same physical state (here the fluid state).  
This free-energy of mixing always comprises a favorable entropic contribution, first introduced by Gibbs. It usually comprises also a contribution associated to the interaction between the different types of molecules which can be either favorable (in a few rare cases) or unfavorable (as a general rule). When unfavorable mixing interactions dominate the entropy of mixing, the homogeneous mixture is no longer stable and separates into two or more phases of different compositions. When a solution is composed of molecules that are similar enough to be substituted for without cost (similar size, shape or chemical properties) the free-energy of mixing reduces to the Gibbs term, and such solutions are said to be \textit{ideal}. For instance in binary mixtures the free-energy of mixing reads $G^{\mathrm{mix}, \mathrm{ideal}} = N k_B T (x\ln(x)+(1-x)\ln(1-x))$ with $N$ the total number of molecules (or equivalently $n$ moles), $k_B$ the Boltzmann constant, $T$ the absolute temperature and $x$ the molar fraction of the first component. 

When the different molecules resemble but are not exactly similar, the entropy of mixing differs from the ideal Gibbs expression. It is then common to account for this difference by introducing a phenomenological quadratic interaction term $N k_B T B x(1-x)$ (again for binary systems, see \textit{e.g.}~\cite{Atkins_PhysicalChemistry}) approach known as theory of \textit{regular solutions}. The parameter $B$ (also commonly denoted $\chi$ in related polymer science theories) captures to leading order the effect of mutual interactions among the two different molecules. A negative $B$ value corresponds to a favorable, promixing trend while a positive $B$ value describes unfavorable mixing interactions. Within this approach an equimolar mixture becomes unstable whenever $B$ exceeds the critical value $B_c=2$. 

The determination of the interaction parameters $B$ is both desirable and challenging~\cite{2009_Almeida}. Knowing $B$ is important when it comes to predicting the phase behavior of lipid mixtures of complex compositions and their critical properties~\cite{2000_Nielsen_Mouritsen, 2008_Honerkampsmith_Keller}. Used in conjunction with a field theory for composition order parameters, it allows to model numerous situation of interest such as the lipid mediated protein interactions or the wetting of membrane inclusions~\cite{1978_Owicki_Mcconnell, 1997_Gil_Ipsen, 1998_Gil_Zuckermann}.  Challenges in the determination of $B$ originate from its very thermodynamic nature. Even though this parameter originates primarily from favorable/unfavorable interactions among neighbors, its quantitative value results from multiple lipid correlations and cannot be obtained simply by restricting the study to a pair of neighboring molecules. This is true already when one reduces lipid mixtures to binary lattice gas problem, where such mean-field predictions are not quantitatively accurate~\cite{1993_Huang_Feigenson, 1993_Huang_Feigenson_2}. This is even more true given the complexity of realistic, coarse-grained  or atomistic representations of the lipids used in modern molecular dynamics systems.  

The purpose of the current manuscript is to determine the effective $B$ parameter associated to a pair of lipids based on molecular dynamics simulations, assuming that the numerical avatars of these lipid molecules are faithful enough to the thermodynamics of mixing of the real systems. Our approach to determining $B$ is based on analysing the composition fluctuations of a simulated binary system, using  di-palmitoyl-phosphatidyl-choline (DPPC) and di-linoleoyl-phosphatidyl-choline (DLiPC) as an example. The chosen force-field was the coarse-grained model SPICA~\cite{2010_Shinoda_Klein,2019_Seo_Shinoda}, running using the parallel LAMMPS code~\cite{1995_Plimpton}. 

\begin{figure}
\begin{tabular}{cc}
\resizebox{0.22\textwidth}{!}{\includegraphics{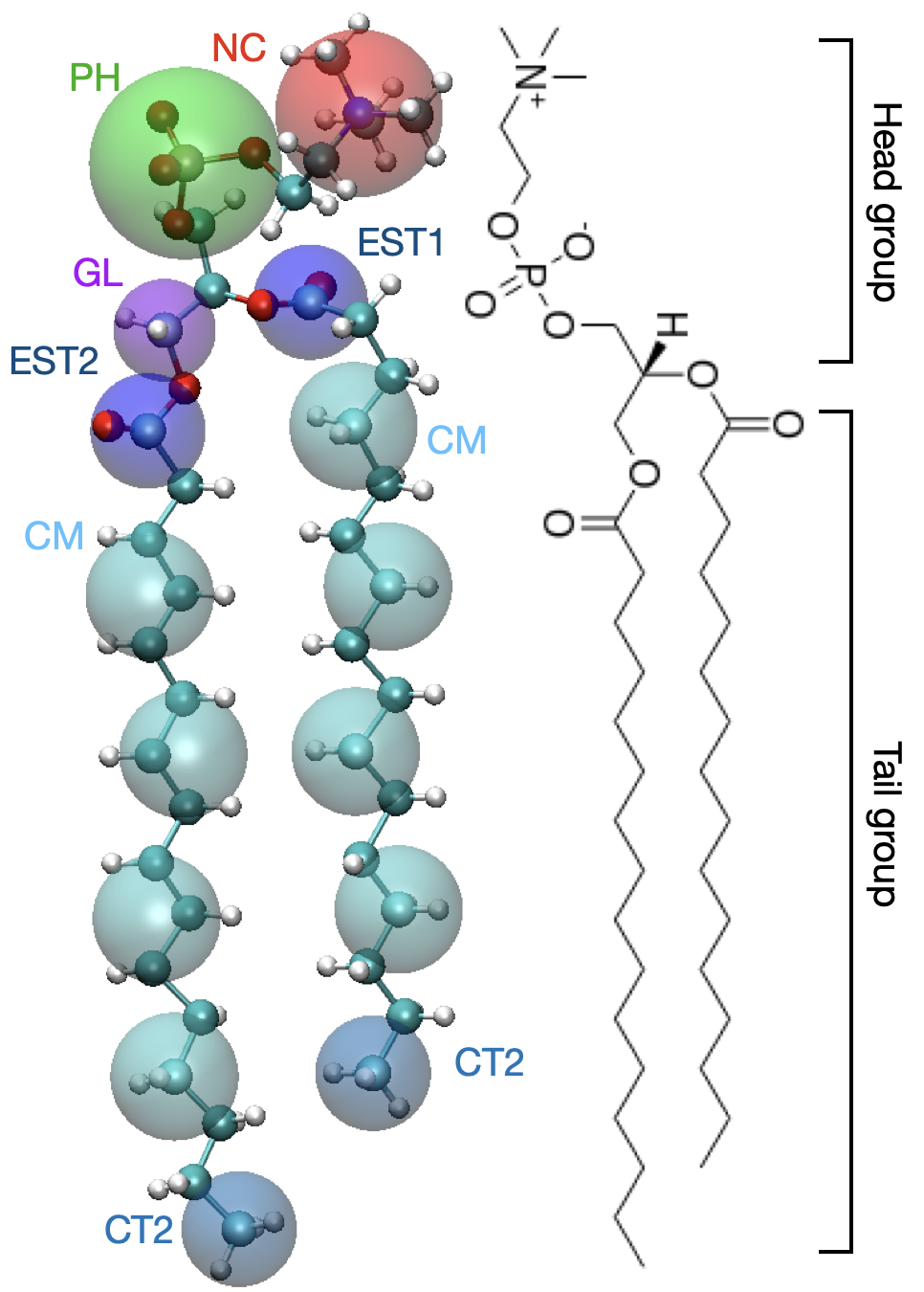}}%
&
\resizebox{0.25\textwidth}{!}
{\includegraphics{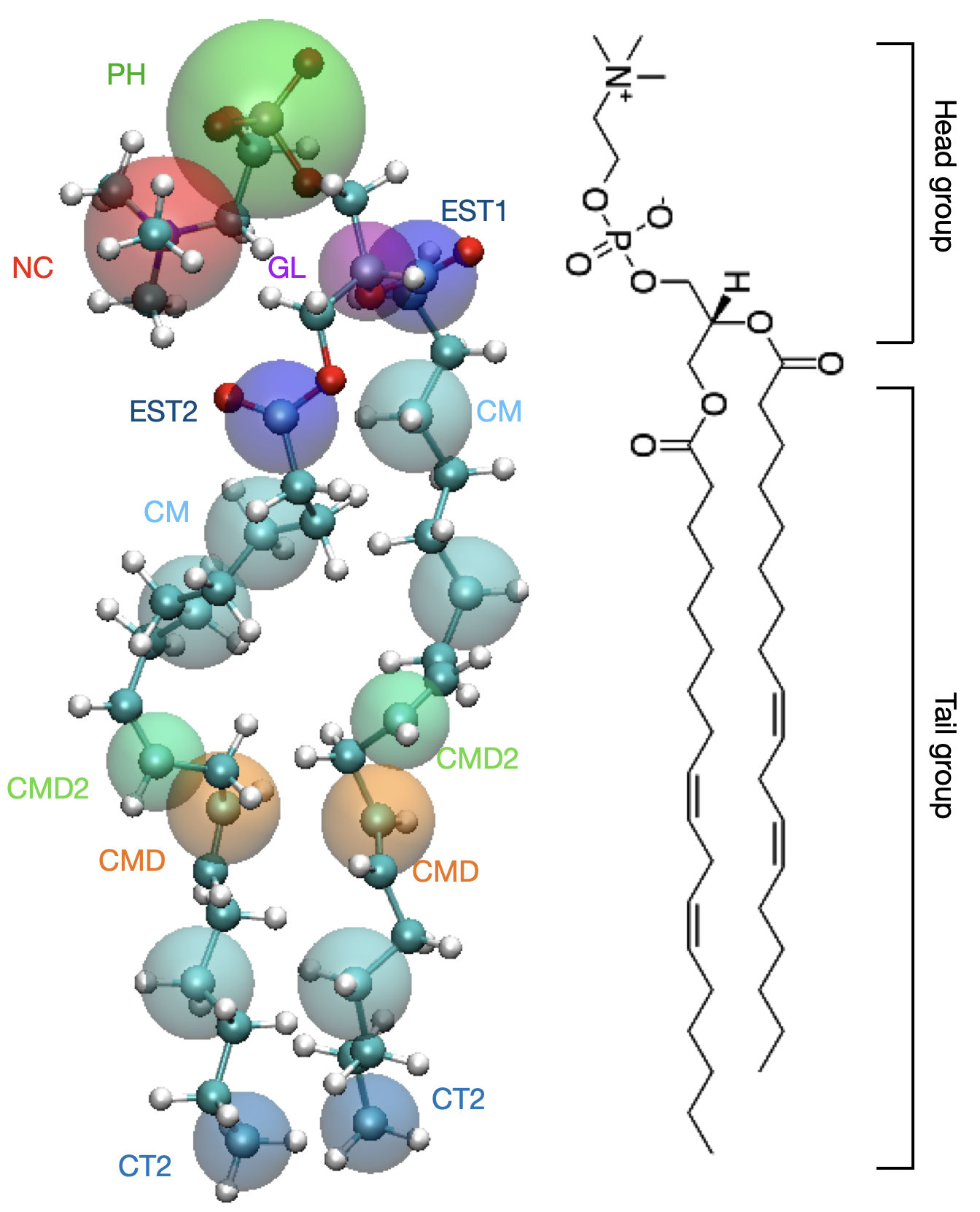}}%
\end{tabular}
\caption{\label{fig:DppcDlipcSpica} SPICA coarse-grained mapping and chemical formula of DPPC (left) and DLiPC (right). }%
\end{figure}


DPPC is a common phospholipid that forms robust bilayers, with a melting transition point at 41$^{\circ}$C; DLiPC has a double unsaturation conferring a much lower main transition temperature ($< 0^{\circ}$C), \textit{i.e.} possesses a much more disordered hydrophobic chain region at room temperature.  The system is simulated at 298K ensuring (this was checked)
an homogeneous fluid state. Both lipid molecules have the same head-group, and differ in the internal organization of their chain-groups. They are therefore similar enough for attempting a description in terms of theory of regular solution. 

Figure~\ref{fig:SnapshotDPPC} presents a snapshot of a pure DPPC bilayer. Figure~\ref{fig:DppcDlipcSpica} shows the detailed chemical composition of both DPPC and DLiPC compounds and their SPICA coarse-grained (CG) representations.

\subsection{Organization of the manuscript}

Our methodology makes several assumptions regarding the statistical thermodynamics of bilayer lipid mixtures and relies on a few non trivial results. Therefore Section~\ref{sec:Theory} of the manuscript is dedicated to introducing the theoretical concepts and approximations underlying the present work. First, a general formalism of density fluctuations is established for 2d fluids and the connection is made with the thermodynamics of mixing of the lipids.  Next we explain how to treat lipids of symmetric leaflets compositions using  binary replication. We then discuss the issue of optimally projecting a real leaflet (or bilayer) onto a plane, reducing 3d lipid molecules to single $(x_{j},y_{j})$ points, and we define 2d structures factors associated to the lipid bilayer.  We end the Section with an analysis of the lipids tilt and bilayer undulations, and derive a useful compressibility relation.  A shorter Section~\ref{sec:Numerics} introduces the model and the numerical simulation conditions. Some results obtained for a pure DPPC bilayer are exposed in Section~\ref{sec:PureBilayer} where they are critically discussed. Section~\ref{sec:Mixture} shows the results obtained for our DPPC-DLiPC pair of lipids and the $B$ values that were obtained. Discussion, methodology and perspectives concludes the work (Section~\ref{sec:Methodology}). A number of appendices provides additional details that would obscure Section~\ref{sec:Theory}.

\medskip
\section{Theory}
\label{sec:Theory}
\subsection{Thermodynamics of two-dimensional regular solutions}

Considering $p$ components indexed with $\alpha=1,\ldots, p$, one defines the free-energy of mixing $G^{\mathrm{mix}}$ as the difference in free-energy between the homogeneous, mixed system and the juxtaposition of the $p$ separated pure components assumed to be in the same physical state (phase).
\begin{equation}
\label{eq:defFreeEnergyMixing}
\begin{split}
    G^{\mathrm{mix}}(T,\sigma,&N_{1},\ldots,N_{p}) =\\  
    &G(T,\sigma, N_{1},\ldots,N_{p} )-\sum_{\alpha} G(T,\sigma,N_{\alpha})
\end{split}
\end{equation}
with $N_{\alpha}$ is the number of molecules $\alpha$, $T$ the temperature, $\sigma$ the surface tension of the bilayer (conjugated to the area) and $G$ the usual notation for the Gibbs free-energy. 
With $N = \sum_{\alpha=1}^{p} N_{\alpha}$ the total number of molecules one introduces the molecular fractions $x_{\alpha}$ and take advantage of the homogeneity of $G$ with respect to its extensive variables $N_{\alpha}$ to write 
\begin{equation}
\label{eq:NotationG}
\begin{split}
   &N = \sum_{\alpha=1}^p N_\alpha \,;\, 
    x_{\alpha} = \frac{N_{\alpha}}{N}; \\
    & G^{\mathrm{mix}}(T,\sigma,\{N_{\alpha}\}) = N G^{\mathrm{mix}}(T,\sigma,\{x_{\alpha}\}).
\end{split}
\end{equation}
The regular solution theory provides a simple approximation of the free-energy of mixing
\begin{equation}
\label{eq:MultiRegularSolution1}
\begin{split}
    \frac{G^{\mathrm{mix}}(T,\sigma,\{N_{\alpha}\})}{k_B T N} = & 
    \sum_{\alpha} x_\alpha\ln(x_{\alpha})\\
    &+\frac{1}{2}\sum_{\alpha,\beta=1}^{p} B_{\alpha\beta} x_\alpha x_\beta
\end{split}
\end{equation}
which for a binary mixture $x_1=x$, $x_2=(1-x)$ reduces to 
\begin{equation}
\begin{split}
    &\frac{G^{\mathrm{mix}}(T,\sigma,N_{1},N_{2})}{k_B T N} = \\
    &\hspace{0.7cm} x\ln(x)+(1-x)\ln(1-x)+Bx(1-x).
    \label{eq:BiRegularSolution}
\end{split}
\end{equation}
The interaction parameter $B$ defined above is dimensionless and expected to be of the order of unity. For vanishing $B$ we recover the expression for the mixing of an ideal solution and $G^{\mathrm{mix}}$ is a convex function of $x$. Upon increasing $B$ it becomes a non-convex function of $x$ ($B_c=2$) prompting a double tangent construction for minimizing the free-energy by separating the system in two optimal stable phases. 

The regular solution model provides expressions for the chemical potentials $\mu_1,\mu_2$ of species $1$ and $2$.
\begin{eqnarray}
    \mu_1(T,\sigma,x)&=& \left.\frac{\partial G}{\partial N_1}\right|_{N_2,T,\sigma}\nonumber\\
    &=& \mu_1^{(0)}(T,\sigma) + k_B T(\ln(x)+B(1-x)^2);\\
    \mu_2(T,\sigma,x)&=& \left.\frac{\partial G}{\partial N_2}\right|_{N_1,T,\sigma}\nonumber\\ 
    &=& \mu_2^{(0)}(T,\sigma) + k_B T(\ln(1-x)+Bx^2).
\label{eq:RegularSolutionChemicalPotential}
\end{eqnarray}
Finally the derivative of the chemical potentials with respect to the other particle number plays an essential role in our approach and reads
\begin{equation}
    \mu_{12} = \left.\frac{\partial^2 G}{\partial N_1\partial N_2}\right|_{N_2,T,\sigma} = \frac{k_B T}{N} \left(-1+2Bx(1-x)\right).
\label{eq:MuCrossDerivative}
\end{equation}

\subsection{Density fluctuations in two-dimensional fluids}

We restrict the discussion to binary $p=2; \alpha=\{1,2\}$ and pure $p=1$ ($\alpha$ omitted) fluids. We consider $N_{\alpha}$ particles in a two dimensional surface parametrized by $(x,y)$ and of area $A$ chosen for convenience as a square of dimensions $L_x=L_y = \sqrt{A}$. Periodic boundary conditions (PBC) are assumed. 

We introduce the number density operators
\begin{equation}
n_{\alpha}(\mathbf{r}) = \sum_{j=1}^{N_{\alpha}} \delta(\mathbf{r}-\mathbf{r}_{\alpha,j})
\end{equation}
with $j$ the particle index, $\mathbf{r}_{\alpha,j}$ the position of the $j$th particle of species $\alpha$ and $\delta(\mathbf{r})=\delta(x)\delta(y)$ the 2d Dirac distribution. The density operators can be expressed in Fourier space
\begin{eqnarray}
    \hat{n}_{\alpha,\mathbf{q}} &=& \int_{\mathcal{S}} \mathrm{d} \mathbf{r} \exp(i\mathbf{q}.\mathbf{r}) n_{\alpha}(\mathbf{r}) \nonumber\\
    &=& \sum_{j=1}^{N_{\alpha}} \exp(i\mathbf{q}\cdot\mathbf{r}_{\alpha,j}).
    \label{eq:FourierCoefficient}
\end{eqnarray}
The observables $\hat{n}_{\alpha,\mathbf{q}}$ can be directly obtained from simulations, provided $\mathbf{q}$ belongs to a set of vectors  $(2\pi n_x/L_x,2\pi n_y/L_y)$ commensurate with the simulation box size $(L_x,L_y)$ and $(n_x,n_y)$ integer numbers. The integration domain $\mathcal{S}$ corresponds to a rectangle $\{0\leq x<L_x;\; 0\leq y < L_y\}$.

We denote with angular brackets $\langle \;\cdot\;\rangle$ the canonical averages of the fluid particles (constant area $A$, temperature $T$ and number of particles $N_{\alpha}$). The $\mathbf{q} = 0$ component of $\hat{n}_{\alpha,\mathbf{q}}=N_{\alpha}$ is a constant quantity. All other values $\mathbf{q} \neq 0$ have vanishing average if the fluid is homogeneous, 
\begin{equation}
\langle \hat{n}_{\alpha,\mathbf{q}}\rangle = 0 
\end{equation}
and therefore the canonical average of products $\hat{n}_{\alpha,\mathbf{q}} \hat{n}_{\beta,-\mathbf{q}}$ correspond to quadratic fluctuations in the number of particles. We have in particular for identical fluid indices $\alpha=\beta=1,2$: 
\begin{equation}
    \langle \hat{n}_{\alpha,\mathbf{q}} \hat{n}_{\alpha,-\mathbf{q}}\rangle = N_{\alpha} + \frac{N_{\alpha}^2}{L_xL_y} \tilde{H}_{\alpha\alpha}(\mathbf{q}),
    \label{eq:DensityFluctuationsDefinitionPure}
\end{equation}
%
while for distinct indices, one has: 
\begin{eqnarray}
    \langle \hat{n}_{\alpha=1,\mathbf{q}} \hat{n}_{\beta=2,-\mathbf{q}}\rangle &=& \frac{N_{1}N_{2}}{L_xL_y} \tilde{H}_{12}(\mathbf{q})\nonumber\\
    &=& \langle\hat{n}_{\alpha=1,-\mathbf{q}} \hat{n}_{\beta=2,\mathbf{q}}\rangle\nonumber\\
    &=& \langle\hat{n}_{\alpha=1,\mathbf{q}} \hat{n}_{\beta=2,-\mathbf{q}}\rangle^{*}.
    \label{eq:DensityFluctuationsDefinitionCross}
\end{eqnarray}
The fact that $\tilde{H}_{12}(\mathbf{q})$ equals its complex conjugate follows from the expected invariance of the system with respect to rotation symmetry $\mathbf{r}_{\alpha,j} \to -\mathbf{r}_{\alpha,j}$.

So far the symmetric matrix $\tilde{H}_{\alpha\beta}$ is just a notation and does not bring anything new to the description of the system. The interest of such quantities appears when one relates it to the statistical description of point-like particles in the continuous thermodynamic limit.  When describing the structure of the fluid it is necessary to introduce the pair correlation functions (radial distribution functions) $g_{\alpha\beta}(r)$~\cite{BenNaim_WaterSolutions,Hansen_McDonald_SimpleLiquids,  Egelstaff_LiquidState}. With $\rho_{\alpha}$ defined as the fluid density $N_{\alpha}/(L_xL_y)$ the pair correlation functions $g_{\alpha\beta}(r)$ expresses that the probability density of finding a particle $\beta$ at distance $r$ from a given particle $\alpha$ is $\rho_{\beta}g_{\alpha\beta}(r)$. This probability is isotropic and $g_{\alpha\beta}(r)=1$ in the absence of interaction (all quantum effects neglected). The presence of position correlations is therefore given by the connected density product $h_{\alpha\beta}(r)=g_{\alpha\beta}(r)-1$. 

The infinite space Fourier transforms of $h_{\alpha\beta}$ is well defined and noted 
\begin{equation}
\tilde{h}_{\alpha\beta}(\mathbf{q}) = \int_{-\infty}^{\infty} \dd x\int_{-\infty}^{\infty}\dd y\, h_{\alpha\beta}(\mathbf{r}) e^{i\mathbf{q}\cdot\mathbf{r}} 
\end{equation}

As the area $A=L_xL_y$ of our finite systems increases, we expect the density correlations $\tilde{H}(\mathbf{q})$ to approach $\tilde{h}(q)$ whenever the reciprocal space vector $\mathbf{q}$ allows it to be computed (commensurate with the system size). In particular $\tilde{H}(q=||\mathbf{q}||)$ becomes isotropic even though the box is rectangular. Then for large systems and moderate values of $q$  
\begin{eqnarray}
    \frac{1}{N_{\alpha}} \langle \hat{n}_{\alpha,\mathbf{q}} \hat{n}_{\alpha,-\mathbf{q}}\rangle & \simeq & 1 + \rho_{\alpha} \tilde{h}_{\alpha\alpha}(q)\nonumber\\
    \frac{L_x L_y}{N_{\alpha}N_{\beta}}\langle \hat{n}_{\alpha,\mathbf{q}} \hat{n}_{\beta,-\mathbf{q}}\rangle &\simeq&   \tilde{h}_{\alpha\beta}.
\end{eqnarray}
In the pure fluid case, $\tilde{h}$ defines also the structure factor of the fluid 
\begin{equation}
S(q) = 1 + \rho\tilde{h}.
\end{equation}

\subsection{The Kirkwood-Buff integrals and how to obtain those quantities in canonical simulations of modest sizes. }

Kirkwood and Buff (KB) have shown that integrals of the form 
\begin{equation}
G_{\alpha\beta} = 
\int_{-\infty}^{\infty}\dd x\int_{-\infty}^{\infty}\dd y\, h_{\alpha\beta}(\mathbf{r}) 
\label{eq:KB-Integrals}
\end{equation}
are related to a number of thermodynamical quantities of interest, including the interaction parameter $B$ when the theory of regular solutions is accurate~\cite{1951_Kirkwood_Buff}. The KB approach generalizes the well-known relation between the structure factor at $q=0$ and the compressibility of the  pure fluid case. 

The Kirkwood-Buff approach is properly formulated in the grand-canonical ensemble, \textit{i.e.} an ensemble where the number of particles $N_{\alpha}$ varies at fixed chemical potential. It is based on the  observation that the grand-canonical fluctuations of number of particles  are on the one hand related  to the derivative of average number of particles with respect to the chemical potential
\begin{equation}
 \frac{\langle N_{\alpha}N_{\beta}\rangle -\langle N_{\alpha}\rangle\langle N_{\beta}\rangle}{k_B T} =\left.\frac{\partial \langle N_{\alpha}\rangle}{\partial \mu_{\beta}}\right|_{A,T} = \left.\frac{\partial \langle N_{\beta}\rangle}{\partial \mu_{\alpha}}\right|_{A,T}
    \label{eq:GrandCanonicalNumberOfParticles}
\end{equation}
and on the other hand to the KB integrals 
\begin{equation}
 \langle N_{\alpha}N_{\beta}\rangle -\langle N_{\alpha}\rangle\langle N_{\beta}\rangle = \langle N_{\alpha}\rangle\delta_{\alpha\beta} +\frac{\langle N_{\alpha}\rangle\langle N_{\beta}\rangle  }{A} G_{\alpha\beta}
 \label{eq:GrandCanonicalKBPure}
\end{equation}
with $\delta_{\alpha\beta}$ the Kronecker delta. 

In the thermodynamic limit one makes no distinction between the observable $N_{\alpha}$ and its average $\langle N_{\alpha}\rangle$. In this limit the Jacobian matrix $(\partial N_{\alpha}/\partial \mu_{\beta})_{A,T}$ is simply the matrix inverse of $(\partial \mu_{\alpha}/\partial N_{\beta})_{A,T}$. As eq.~(\ref{eq:MuCrossDerivative}) shows, there is a connection between the regular solution parameter $B$ and the derivative 
$\mu_{12} = (\partial \mu_{1}/\partial N_{2})_{\sigma,T}$.

Proceeding along these lines and leaving technicalities to the appendices we now provide the essential relations that are needed in this manuscript. Notations are borrowed from Ben-Naim~\cite{BenNaim_WaterSolutions}. 

\begin{itemize}
    
\item Pure case 
\begin{equation}
     1+\rho G_{11} = \rho k_B T \chi_T.
     \label{eq:compressibilityPureCase}
\end{equation}
where $\chi_T$ stands for the isothermal area compressibility of the system. 

\item Binary mixture
\begin{equation}
\begin{split}
&\Delta = G_{11}+G_{22} -2G_{12};\\
&\eta = \rho_1+\rho_2 +\rho_1\rho_2\Delta;\\
&\zeta = 1+\rho_1 G_{11}+\rho_2 G_{22}\\
&\phantom{\zeta = 1+\rho_1 G_{11}}+\rho_1\rho_2(G_{11}G_{22}-G_{12}^2);\\
&\mu_{12} = -\frac{k_B T}{A\eta};\\
&\chi_T = \frac{\zeta}{k_B T\eta}.
\end{split}
\label{eq:KBRelationsBinaryMixtures}
\end{equation}
\end{itemize}

A further look at eq.~(\ref{eq:MuCrossDerivative}) allows us to derive the relation between $B$ and the KB integrals. 
\begin{equation}
    B = \frac{\rho}{2}\left[\frac{\Delta}{1+\rho x(1-x) \Delta}\right].
    \label{eq:KBEquationForB}
\end{equation}
It is important to stress that the expression for $\mu_{12}$ in eq.~(\ref{eq:KBRelationsBinaryMixtures}) has a \textit{general validity} while expression~(\ref{eq:KBEquationForB}) makes assumption on the expression of $G^{\mathrm{mix}}$ and its validity is \textit{limited to the validity of the regular solutions approximation}. 

Should another model for $G^{\mathrm{mix}}$ be considered that expression~\ref{eq:KBEquationForB} would not be valid. Connection with such a model should be done independently by means of the quantity $\mu_{12}$.

\subsection{Specific features in membrane thermodynamics}

Lipid bilayers are very thin 3D films. The lipid bilayer volume is thus conjugated to a 3D isotropic solvent pressure $P$. In addition, a self-assembled lipid bilayer can stand a low but finite tension $\sigma$. In the case of a positive tension, the membrane is in a metastable state which can eventually lead to an activated pore opening followed by the rupture of the film~\cite{2003_Evans_Rawicz}. The opposite case of negative tension leads to a buckling instability beyond a certain threshold. Fortunately in practice moderate values of the surface tension are compatible with a long lived quasi-equilibrium state, and both pressure $P$ and tension $\sigma$ can be taken as independent intensive thermodynamic control parameters.  

Lipid hydration poses another question. It is established \cite{Marsh_HandbookLipidBilayers, Marsh_HandbookLipidBilayers2, Cevc_Marsh_PhospholipidBilayers, Evans_Wennerstrom_ColloidalDomain}  that the membrane state depends on hydration, \textit{i.e.} the ratio of water molecules per lipid available in the solution. In many cases however water is in large excess as compared to the lipid molecules present. It becomes possible to consider the self-assembled lipid film as a phase coexisting with an almost pure water solution phase. This is because the molecular phospholipid solubility is extremely low. When focusing only on membrane thermodynamics, the water solution plays the role of a reservoir of constant chemical potential. Even though lipids are strongly hydrated, water molecules play only an implicit role in the lipid interactions. The additional chemical component (water) compensates the additional phase (water solution) in the Gibbs phase rule counting. 

We therefore identified the following relevant thermodynamic variables in our problem: temperature $T$, isotropic solvent pressure $P$, membrane tension $\sigma$, membrane area $A$, lipid number components $N_{\alpha}$. These thermodynamic parameters are all specified in the molecular dynamics simulations. These simulations also specify the number of water molecules $N_w$ which is shown to have little influence on the simulation results as long as it is large enough, in agreement with the two phase coexistence arguments stated above. Finally, the isotropic pressure plays only a marginal role until it reaches values of the order of 10-100 bars, because both water solution and lipid membranes are little compressible condensed states. 

Further considerations on the number and role of thermodynamics parameters necessary to properly describe a membrane vesicle can be found in the work of Diamant~\cite{2011_Diamant}.

\subsection{Density fluctuations in the hydrodynamic regime}

It remains to estimate the KB integrals from finite size simulations. The pair correlation functions of simple fluids are well defined and do not depend on the ensemble in the thermodynamic limit. We therefore expect a regular $q\to 0$ limit to the $\tilde{h}_{\alpha\beta}(q)$ correlations:
\begin{equation}
\lim_{q \to 0} \tilde{h}_{\alpha\beta}(q) = \tilde{h}_{\alpha\beta}(0) = G_{\alpha\beta}.
\label{eq:KBInPractice}
\end{equation}
As our finite size density correlation functions $\tilde{H}_{\alpha\beta}(\mathbf{q})$ approximate $\tilde{h}_{\alpha\beta}(q)$ for commensurate nonvanishing wave-vectors $\mathbf{q} = (2\pi m_x/L_x, 2\pi m_y/L_y)$ ($m_x,m_y$ taking integer values) one is reduced to the problem of \textit{extrapolating} $\tilde{h}_{\alpha\beta}(0)$ from a sequence of small $q$ values. 

For simple 2d fluids, one has in principle the freedom to choose $L_x$ and $L_y$ as large as wanted. As a result the smallest $q$ value $2\pi/L_x$ can be taken as close to 0 as wanted. When it comes to numerical simulations there are however serious obstacles to overcome. As the size of the system increases, the computational cost fast become unbearable.  Even with access granted  to large parallel computing facilities, it takes longer and longer to obtain satisfactory statistical estimates of the correlation products $\tilde{H}_{\alpha\beta}(\mathbf{q})$. This is because the characteristic relaxation time of a fluctuation mode $\hat{n}_{\mathbf{q}}(t)$ diverges like $q^{-2}$ at small $q$. The time needed to extract statistically significant information from noise fast becomes prohibitive. In practice there are limits to the sizes of the simulation box that can be simulated while maintaining low statistical errors in $\tilde{H}_{\alpha\beta}(\mathbf{q})$.

The case of lipid bilayers brings even more stringent restrictions. On the one hand, as will be explained below, reducing lipid molecules to a single point coordinates result in dealing with quite low numbers of effective particles $N_{\alpha},N_{\beta}$ and poor statistics. On the other hand, by enlarging the system one favors Helfrich undulations and departs from the idealized picture of a 2d flat simple fluid. One therefore must accept a compromise between tractable computation times, quality of the statistics and low bilayer roughness. 

The low $q$ density fluctuations are called hydrodynamic modes. They enable us to probe the many-body thermodynamical properties of the fluid. As soon as $q\sqrt{a_l} \ll 1$ with $a_l$ the area per molecule in the fluid, it becomes possible to replace the density operator $n_{\alpha}(t)$ by a slowly varying coarse-grained density field $\rho_{\alpha}(\mathbf{r},t)$ such that 
\begin{equation}
\begin{split}
\hat{n}_{\alpha,\mathbf{q}}(t) = \int_0^{L_x} \dd x\, &\int_0^{L_y} \dd y\, 
[\cos(q_x x+q_y y)\\
& +i \sin(q_x x+q_y y)] \rho_{\alpha}(\mathbf{r},t)
\end{split}
\label{eq:FourierTransformCGField}
\end{equation}
and conversely the instantaneous density can be represented as a Fourier series 
\begin{equation}
\begin{split}
\hat{n}_{\alpha}(\mathbf{r},t) =  \sum_{m_x,m_y}
[\cos(q_x x+q_y y)- & i \sin(q_x x+q_y y)] \\
& \times\hat{n}_{\alpha,q_x,q_y}(t)
\end{split}
\end{equation}
where $(q_x,q_y)=(2\pi m_x/L_x, 2\pi m_y/L_y)$. It is clear that none of the finite $q$ mode changes the total number $N_{\alpha}$ of particles. These modes are free to fluctuate in the canonical ensemble and have the same statistics as the grand-canonical density modes with same $q$. The grand-canonical fluctuations in the total number of molecules are expected to be largely uncoupled to the finite $q$ density fluctuations. This is how a canonical simulation can be used to extrapolate the grand-canonical number fluctuations in the $q\to 0$ limit. 

We believe the described approach to be simpler and more accurate than the one consisting in painfully computing pair correlation histograms and then integrating them spatially with the aim of approximating $G_{\alpha\beta}$. The \textit{structure factor way} that we use requires only the computation and the average of products of 1-body observables $\hat{n}_{\alpha,\mathbf{q}}(t)$.


\subsection{Replication of a symmetric bilayer}

\begin{figure}
\resizebox{0.45\textwidth}{!}{\includegraphics{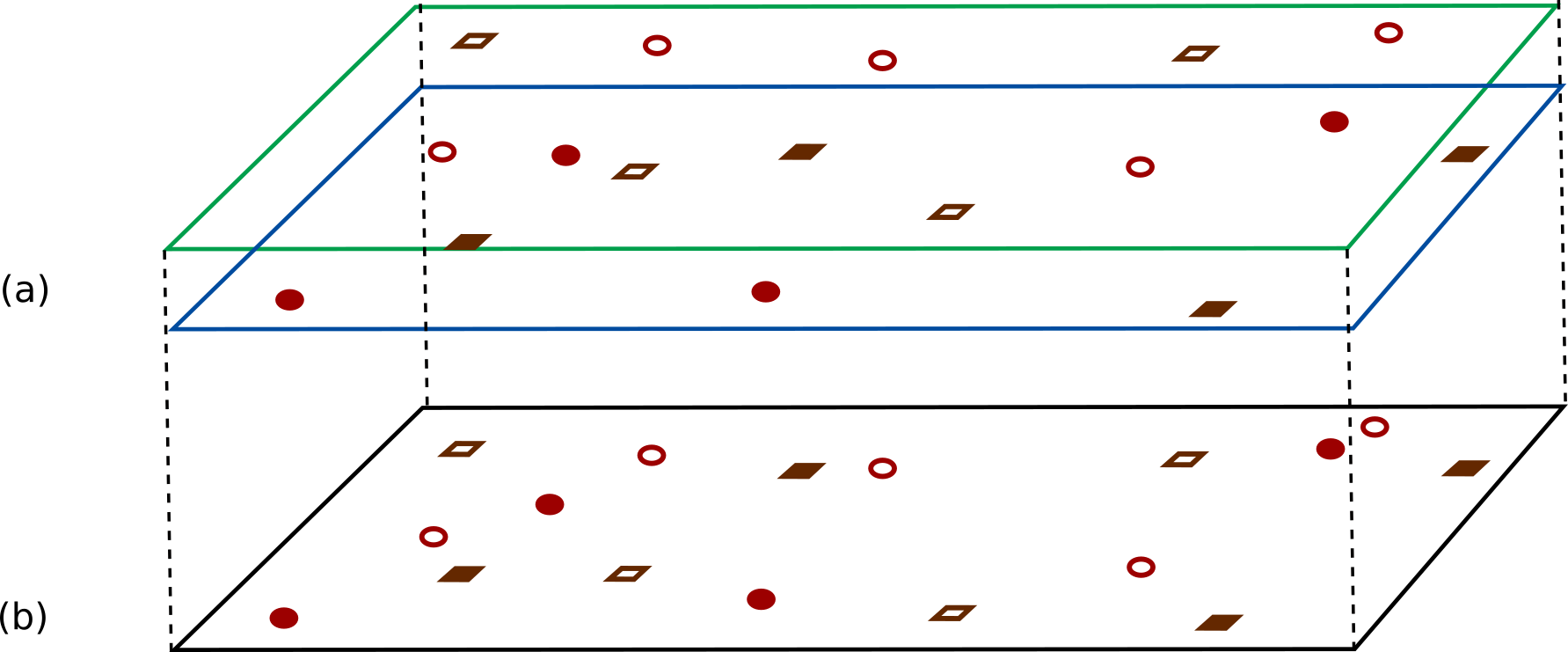}}%
\caption{\label{fig:Replication} 
 This picture illustrates the idea of replication of the bilayer fluid. Each leaflet (blue and green) contains two types of molecules (squares, circles). Molecules from the upper leaflet (open symbols) interact weakly with molecules from the lower leaflet (full symbols) (a). The projection of the system onto a single plane is equivalent to a quaternary lipid mixture, each species possessing a chemical index (square, circle) and a replica, or leaflet index (empty, full) (b). }%
\end{figure}

Lipid bilayers comprise two leaflets. Interleaflet lipid interactions are  weaker than intraleaflet correlations but not necessarily negligible and even when reducing leaflets to flat 2d fluids it remains to account for the positional correlations of lipids located in opposing sides. Formally this can be done by considering lipids belonging to different leaflets as different species thus doubling the number $p$ of molecules. When the bilayer is symmetric, the situation greatly  simplifies and reduces to a particular case of replicated fluid theory, where multiple interacting copies of the same fluids are considered, usually in order to address the consequence of quenched disorder~\cite{1994_Menon_Dasgupta,1995_Pitard_Tarjus, 1996_Mezard_Parisi}. The replicated fluid  approach serves here as a guide to explicit the structure of the double fluid. 

We restrict the discussion to $p=2$ binary mixtures. In addition to the $\alpha$ label, we introduce a leaflet (replica) label $\lambda = 1$ (upper leaflet) and 2 (lower leaflet), see Figure~\ref{fig:Replication}. The $2p$ effective mixture is now labeled by a multi-index $\alpha\to (\alpha,\lambda)$. The correlation functions are labeled by 2 pairs of indices, for instance $\tilde{h}_{\alpha\beta,\lambda\nu}(\mathrm{q}) = \langle\hat{n}_{(\alpha,\lambda),\mathbf{q}}(t)\hat{n}_{(\beta,\nu),-\mathbf{q}}(t) \rangle$.

Replica symmetry means that correlations reduces to the intra- ($\tilde{h}$) and inter- leaflet ($\tilde{h}'$) case.
\begin{equation}
\tilde{h}_{\alpha\beta,\lambda\nu}(\mathbf{q})=
 \delta_{\lambda\nu} \tilde{h}_{\alpha\beta}(\mathbf{q})+ (1-\delta_{\lambda\nu}) \tilde{h}'_{\alpha\beta}(\mathbf{q}).
\end{equation}
Strictly non-interacting fluids corresponds to vanishing $\tilde{h}'_{\alpha\beta}$. 
This new correlation function can be obtained numerically by correlating density fluctuation modes from opposite leaflets, and measures the effect of lipid interactions across the bilayer plane. In the presence of non-vanishing correlations we should treat the system as a 4 components mixture and derive the KB expressions accordingly. It can be shown that when $\tilde{h}'=0$ the block diagonal structure of the density fluctuation matrix leads to a natural generalization of the KB expressions obtained in the binary case. In particular the compressibility of the bilayer is half the compressibility of the monolayer. 

If one denotes by $N_{\alpha}$ the number of $\alpha$ molecules in one replica (leaflet) and $A$ the common area of the two fluids, then 

\begin{equation}
\begin{split}
    \langle(\hat{n}_{\alpha,1,\mathbf{q}}+&\hat{n}_{\alpha,2,\mathbf{q}})(\hat{n}_{\beta,1,\mathbf{q}}+\hat{n}_{\beta,2,\mathbf{q}})\rangle =\\
    & 2N_{\alpha}\delta_{\alpha\beta} +\frac{(2N_{\alpha})(2N_{\beta})}{A}\frac{\tilde{h}_{\alpha\beta}(\mathbf{q})+\tilde{h}'_{\alpha\beta}(\mathbf{q})}{2} 
    \end{split}
    \label{eq:BunchedReplicatedDensities}
\end{equation}
in the left hand side, molecules pertaining to opposite leaflets have been bunched together. In the absence of inter-leaflet correlations eq.~(\ref{eq:BunchedReplicatedDensities}) demonstrates that the intra-leaflet correlations $\tilde{h}_{\alpha\beta}$ can also be obtained by computing and scaling appropriately the correlations of the full bilayer density operators $(\hat{n}_{\alpha,1,\mathbf{q}} + \hat{n}_{\alpha,2,\mathbf{q}})(t)$.  

\subsection{Reducing lipid molecules to point-like objects}

Fluids of point-like molecules interacting  through pair-wise additive interactions are known as simple fluids. Their theory is built on firm ground\cite{Hansen_McDonald_SimpleLiquids,Egelstaff_LiquidState}. As molecules grow in mass, the theoretical description becomes harder and requires approximations based on physical intuition and is specific to each situation (lipids, polymers, amphiphiles, etc.). The grand canonical ensemble fast becomes unpractical in numerical simulations, as addition and removal of macromolecules takes place only with exponentially low acceptance. Kirkwood-Buff relations are nevertheless expected to be valid, describing simple fluids as well as complex macromolecular assemblies. In order to use relations  \ref{eq:DensityFluctuationsDefinitionPure},  
\ref{eq:DensityFluctuationsDefinitionCross}, 
\ref{eq:KBRelationsBinaryMixtures} and 
\ref{eq:KBInPractice} we need an operatorial definition of the density modes $\hat{n}_{\alpha,\mathbf{q}}(t)$. In practice one must determine the pair of coordinates $(x_{\alpha,\lambda},y_{\alpha,\lambda})$ that best localizes each molecule $(\alpha,\lambda)$. 

A simple possibility consists in choosing a representative CG beads in the lipid model (Fig~\ref{fig:DppcDlipcSpica}) dropping its vertical $z_j$ coordinate. A typical SPICA lipid has \textit{ca}~15 distinct constitutive beads leading to as many different possible 2d density definitions. Other choices include the center of mass (\textit{com}, weighted by beads masses) and the pseudo-center of mass (\textit{pcom} obtained from the 15 beads with equal weight per bead).  The $\tilde{h}_{\alpha\beta}(q)$ structures depend on the chosen representation, but their $q\to 0$ limit $\tilde{H}_{\alpha\beta} \equiv G_{\alpha\beta}$ is insensitive to it in the case of a flat system in the thermodynamic limit. The independence of the representation of the KB integrals is for instance discussed by Koga and Widom~\cite{2013_Koga_Widom}. Indeed, one convinces oneself easily that if two representative points of a same molecules are bound by an effective potential of finite range, the difference between the two structures $\tilde{h}_{\alpha\beta}$ is bound by a quadratic term $C q^2$, vanishing in the $q\to 0$ limit.

As numerical simulations restrict ourselves to finite $q_{\min}\sim 2\pi/\sqrt{A}$ values, the $q\to 0$ extrapolation remains dependent on the choice of the representation. In order to mitigate the effect, it is important to compare the different possible structures obtained and find out the \textit{best} representation of the lipid structure. It turns out that the best representation of a lipid depends on the $q$ range considered (Section~\ref{sec:Methodology}). 

The $q$ independence of the $\tilde{H}_{\alpha\beta}$ which can be proven in the thermodynamic limit for flat systems unfortunately breaks down for rough undulating systems, such as thermal lipid bilayers. There is then the need of a deeper analysis and a replacement for relation~(\ref{eq:KBRelationsBinaryMixtures}). 

\subsection{Influence of lipid tilt and membrane undulations on the determination of the KB limit values}

\begin{figure}
\resizebox{0.45\textwidth}{!}{\includegraphics{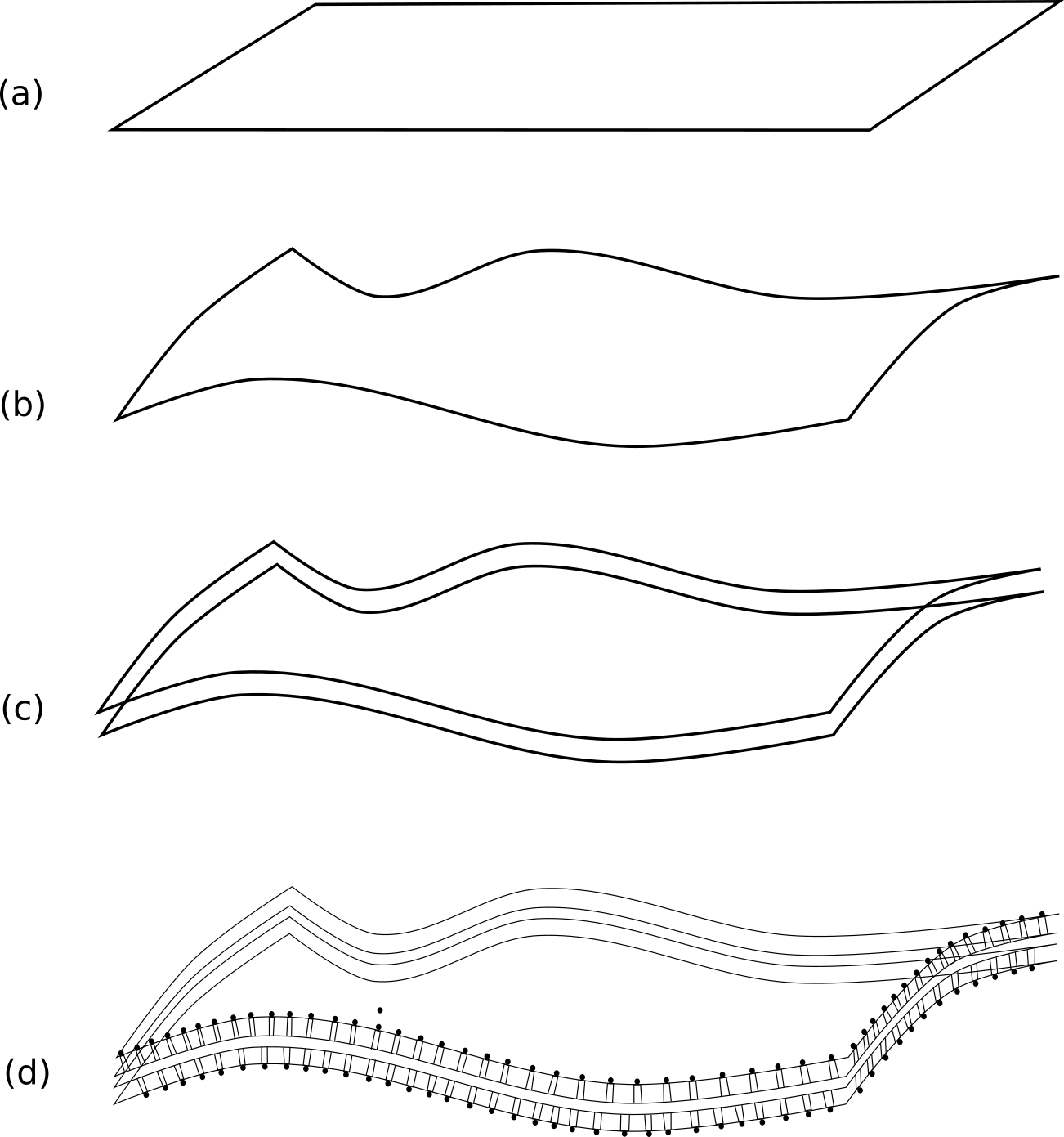}}%
\caption{\label{fig:MembraneModels} The simplest membrane model is the infinitely thin  flat surface (a). Thermal fluctuations distorts the plane which becomes a curved surface described by the Helfrich elastic energy (b). The next level distinguishes the two leaflets introducing new elastic terms such as the area difference elasticity~\protect\cite{1997_Seifert} (c). A bilayer can finally be described with fluid elasto-nematic model~\protect\cite{2011_Watson_Brown} (d).
  }%
\end{figure}

\begin{figure}
\resizebox{0.45\textwidth}{!}{\includegraphics{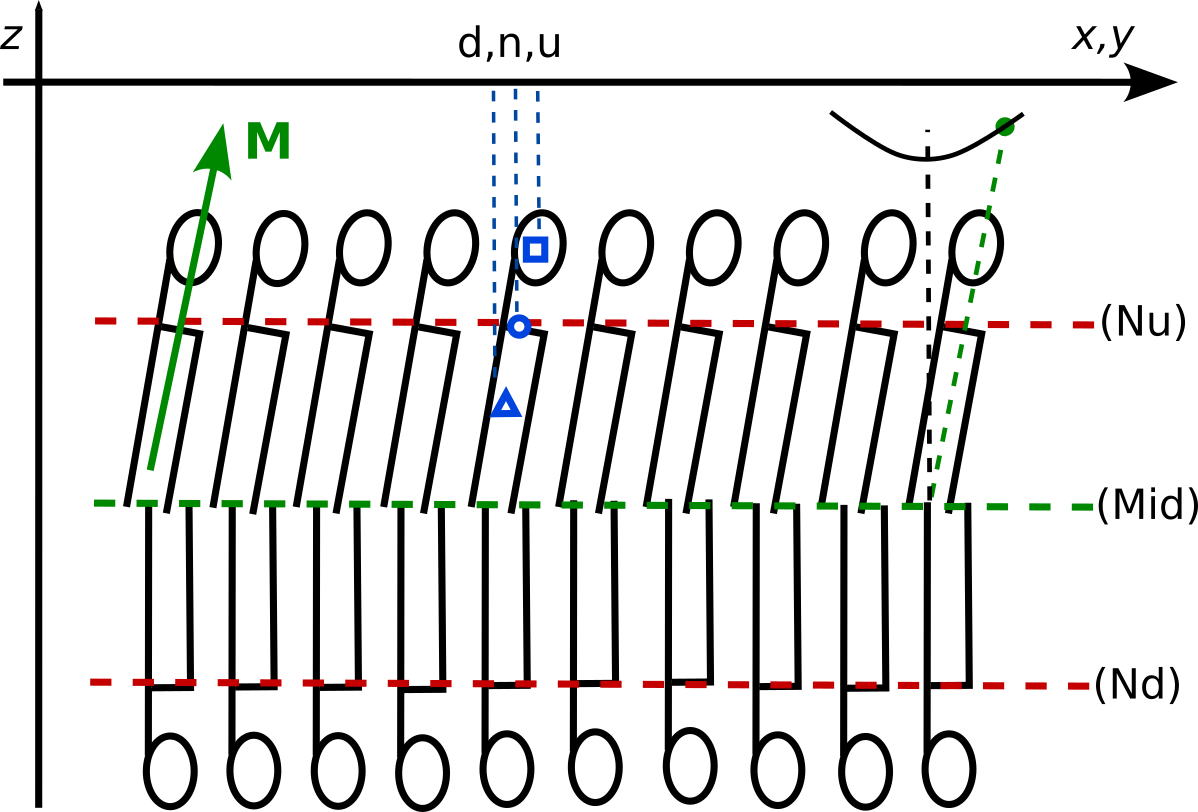}}%
\caption{\label{fig:IllustrationTilt} Illustration of the lipid tilt in flat bilayers. $\mathbf{M}$ is the director vector which is supposed to vary in space on a hydrodynamic scale much larger than the typical lipid separation $\sqrt{a_l}$. Each leaflet tilts independently from the other and has its own neutral surface (Nu for upper leaftet and Nd for down leaflet). The horizontal projection depends on the representative bead chosen (u, blue square: headgroup sitting above the neutral surface, n, blue circle: close to the neutral surface, d, blue triangle: below the neutral surface). The bilayer mid-plane, Mid, is shown with a green dashed line.  }%
\end{figure}

Bilayer membranes can be described by a sequence of models with increasing realism and complexity. The two-dimensional (2d) flat fluid constitutes the most basic description and was the focus of the above sections~(Fig.~\ref{fig:MembraneModels}).

The next step towards realistic membranes is due to Canham and Helfrich who considered a two dimensional non planar surface (2d submanifold) embedded in a three dimensional solvent, subject to a bending elastic energy while displaying lipid fluidity.
In the case of a nearly flat membrane of symmetric leaflet composition, the Canham-Helfrich elastic model depends on two intensive parameters: the bending modulus $\kappa$ and the tension $\sigma$~\cite{1973_Helfrich_2, Safran_Surfaces, Heimburg_BiophysicsMembrane}. 
The next step consists in recognizing that bilayers are made of two leaflets separated by a nanometric distance $\mathcal{D}$ of the order of the membrane thickness (1.5-2.5~nm). 
At this level of description, it is possible to consider a membrane of asymmetric composition with new terms in the elastic energy. Finally, one can deal with the anisotropic lipid molecules with order analogue to nematic liquid~\cite{1974_Marcelja_2, 1980_Priest}. In the latest picture, the anisotropic character of the lipid assembly can be accounted for as a first approximation by a molecular director vector $\mathbf{M}$~\cite{2008_Brown, 2011_Watson_Brown, 2012_Watson_Brown}. Deviation of the molecular orientation of lipid molecules with respect to the bilayer normal will be referred below as \textit{lipid tilt}.

\begin{figure}[t]
\resizebox{0.45\textwidth}{!}{\includegraphics{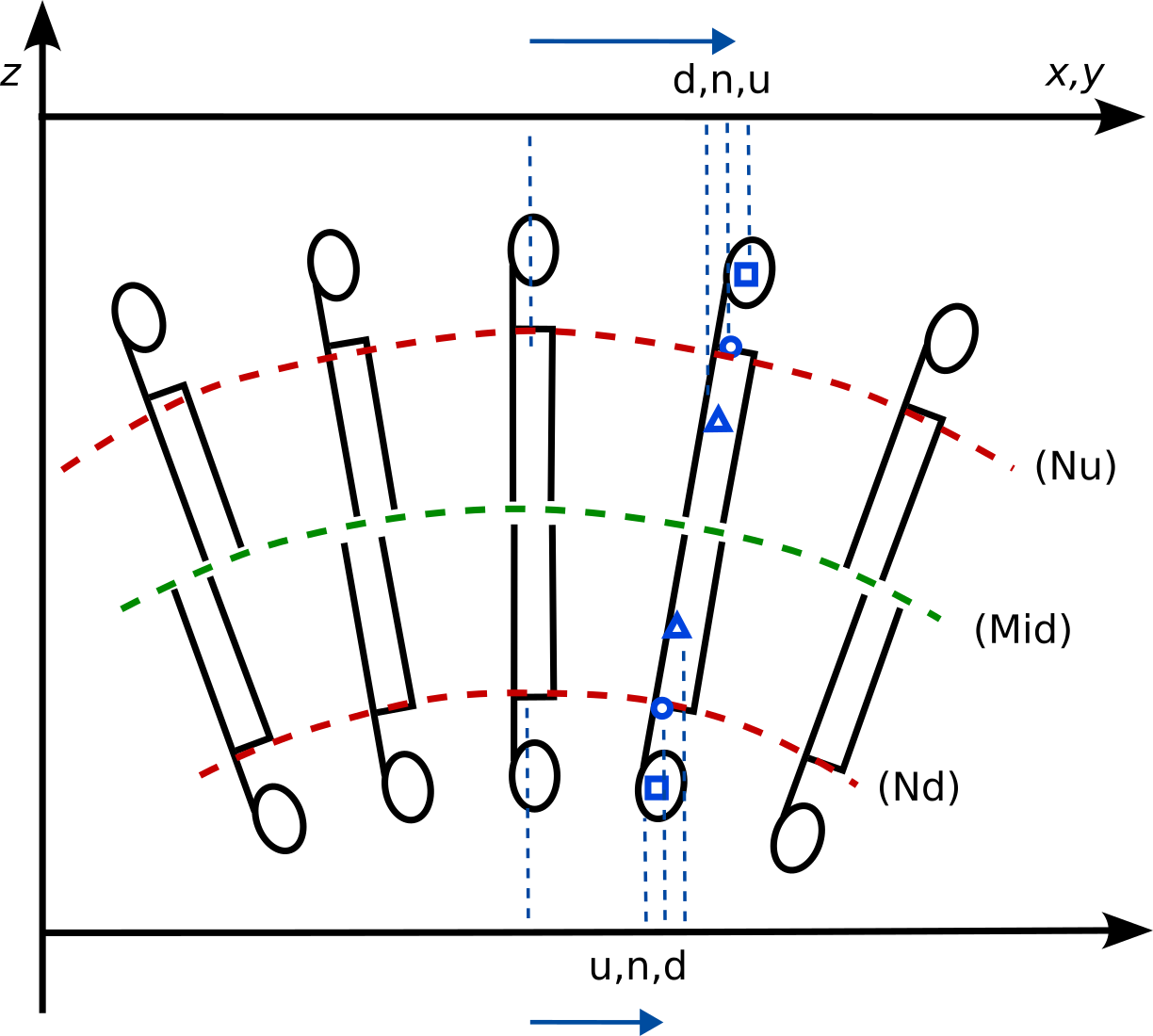}}%
\caption{\label{fig:IllustrationUndulation} 
Schematic illustration of a cylindrically curved membrane, showing a curved mid-plane (Mid, green dashed line). There is no lipid tilt as the director vector $\mathbf{M}$ is everywhere orthogonal to the mid-bilayer surface. The two red dashed planes (Nd, Nu) represent respectively the neutral surfaces of the upper and lower leaflets  around which the lipid molecules are expected to tilt. The upper leaflet is expanded while the lower leaflet is compressed as the result of the curvature  of the bilayer. The relative displacements of the horizontal projections of the lipid depends on the reference bead (same convention as in Figure~\protect\ref{fig:IllustrationTilt}). Note that the ordering of the projections is reversed in the opposite leaflet.   }%
\end{figure}

As shown in Figure~\ref{fig:IllustrationTilt} the horizontal projection of the lipid representative point depends on its tilt angle and on its position along the lipid main axis. We assume for each leaflet the existence of a neutral surface around which lipid molecules can pivot (tilt motion). Intuitively, one understands that lipids with representative points located far above or below the neutral surface will give rise to largest horizontal density fluctuations as compared to lipids with representative points close to the neutral surface.  We therefore assume that there exists an ideal locator of the lipid molecule sitting at the intersection of the lipid axis and the neutral line (blue circle symbol in Figure~\ref{fig:IllustrationTilt}). If a representative bead is chosen further apart on the lipid axis, at a distance $\mathcal{D}$ from the neutral surface, the projections  $\mathbf{r}_{\alpha,j}^{(n)}$  of the best locator and the actual representative bead $\mathbf{r}_{\alpha,j}$ are separated by an horizontal displacement 
$\mathbf{u}_{\alpha,j} \simeq \mathcal{D}\mathbf{M}_{\parallel}$, where $\mathbf{M}_{\parallel}$ stands for the horizontal projection of the director vector.

The tilt displacement contributes to the effective density mode fluctuation by an amount $-\rho_{0}\mathrm{div}(\mathbf{u})$. If one considers now that the best locator is representative for the true density fluctuations, and the true structure factor $S_{\mathrm{best}}(q)$ of the ideal 2d fluid, one finds that the apparent and the true structure factors are related by the expression (pure systems):
\begin{equation}
    NS(q) = NS_{\mathrm{best}}(q) + \rho_0^2 \mathcal{D}^2 \mathrm{q}^2\langle|\mathbf{M_{\parallel}}(\mathbf{q})|^2\rangle.
\end{equation}
The effect of the tilt modes is to add independent fluctuations to the one arising from the finite compressibility of the fluid. In other words, lipid tilt makes the lipid bilayer softer in appearance, more compressible.

It is possible to quantify the thermal fluctuations of the director $\mathbf{M_{\parallel}}$.  We observe that lipid molecules tend to occupy a vertical orientation as if there is a spring connected to the end of  $\mathbf{M_{\parallel}}$ causing a quadratic energy $K_0\mathbf{M_{\parallel}}^2/2$~\footnote{The fluid lipid state is not expect to display any permanent tilt with respect to the normal bilayer unlike the so-called $L_{\beta'}$ gel state for instance}. A thermal equipartition model leads to the expression 
\begin{equation}
    S(q) = S_{\mathrm{best}}(q) +\rho_0^2 \mathcal{D}^2 k_B T \frac{q^2}{K_0}.
\end{equation}
This expression again contributes to a quadratic $q$ deviation from the extrapolated compressibility $S_{\mathrm{best}}(0)$. Tilt modes alone do not prevent from extrapolating the structure factor and the correlations $\tilde{h}_{\alpha\beta}$ to the origin. 

Bilayer out-of-plane fluctuations can be analysed in a similar way. We make the choice of decoupling the bending modes from the tilt modes. This is arbitrary because both lipid tilt and bending are controlled by the same elastic coefficient, as discussed by Watson et al.~\cite{2012_Watson_Brown}. We nevertheless find it simpler to analyse the current situation in this way.

Figure~\ref{fig:IllustrationUndulation} represents a bent lipid membrane, where the leaflets are uncoupled and  lipid tilt is absent. The director vector $\mathbf{M}$ is everywhere orthogonal to the bilayer mid-plane surface. Again, only long wave-vector lipid tilt modulations are considered. With the same notations and conventions, there exists a neutral surface around which the lipid molecules pivot. A best locator (blue circle symbol) is assumed to be representative of the true density fluctuations ($S_{\mathrm{best}}(q)$ in the pure case). Lipids from opposite leaflets are free to slide relative to each other (only thermal equilibrium fluctuations are considered so that interleaflet friction does not play any significant role). 

The neutral surfaces have roughly the same curvature as the mid-bilayer surface. Noting $z(\mathbf{r})$ the vertical  elevation of the neutral surface, they contribute to a relative increase of density caused by a geometrical projection factor $\delta \rho/\rho_0 =(\nabla z)^2/2$ ~\cite{2005_Reister_Seifert}. One can show that this geometrical effect is negligible and does not contribute significantly to the apparent increase in compressibility of the leaflet.

It is clear from the Figure~\ref{fig:IllustrationUndulation} that choosing a representative bead at a distance $\mathcal{D}$ above or below the neutral surface has the effect of increasing the apparent membrane compressibility. The reasoning is similar to the lipid case, with the bilayer normal direction $\mathbf{n}$ playing the role of the lipid director. We have in particular 
\begin{equation} 
\mathrm{div}(\mathcal{D}\mathbf{n}_{\parallel}) = -\mathcal{D}\Delta z
\end{equation}
leading to the relation 
\begin{equation}
    NS(\mathbf{q}) = NS_{\mathrm{best}}(\mathbf{q}) + \rho_0^2 \mathcal{D}^2 \mathbf{q}^4 \langle|\tilde z_{\mathbf{q}}|^2\rangle.
\end{equation}
The average  $\langle|\tilde z_{\mathbf{q}}|^2\rangle$ is  known as the Helfrich spectrum and in the absence of surface tension reads $k_BTL_xL_y/(\kappa q^4)$ with $\kappa$ the mean curvature bending modulus. Due to the compensation between the $q^4$ terms we obtain now the fluctuation expression 
\begin{eqnarray}
    S(\mathbf{q}) &=& S_{\mathrm{best}}(\mathbf{q}) + \rho_0 \mathcal{D}^2 \frac{k_B T}{\kappa}\nonumber\\
     &=& \rho_0 \frac{k_B T}{K_A}  + \rho_0 \mathcal{D}^2 \frac{k_B T}{\kappa}.
    \label{eq:AllFluctuations}
\end{eqnarray}
Expression~\ref{eq:AllFluctuations} is one of our main results. It expresses that the undulations contribute to the effective compressibility by a finite amount. It shows that the extrapolated $q=0$ value of the correlations $\tilde{h}_{\alpha\beta}$ does depend on the choice of the representative bead. 

We make use of eq.~(\ref{eq:AllFluctuations}) to rationalize our results in Section~\ref{sec:Methodology}. 
First, we observe on Figure~\ref{fig:IllustrationUndulation} that because the upper and lower leaflets have opposite trend under bending, combining both leaflets into a single system as in eq.~(\ref{eq:BunchedReplicatedDensities}) should have the effect of canceling, or strongly reducing the influence of the undulations on the apparent increase of compressibility. This is confirmed by simulations.

Second, tilt and undulations affect all lipids in a quasi-similar way and should not be sensitive to the lipid index $\alpha$. Combinations of correlations such as $\Delta$ in eq.~(\ref{eq:KBRelationsBinaryMixtures})
are less sensitive to those than separated terms $\tilde{H}_{\alpha\beta}$. Simulations also confirm that the interaction parameter $B$ though difficult to obtain, does not seem to depend significantly on the choice of the representative bead.

\medskip
\section{Numerics}
\label{sec:Numerics}
\subsection{The SPICA force-field}

The molecular dynamics model used in the present approach is the coarse-grained SPICA force field~\cite{2010_Shinoda_Klein, 2019_Seo_Shinoda}. 
The SPICA coarse-graining maps 3~atoms (hydrogens not comprised) to 1~bead, uses realistic masses and implements long-range electrostatics. The water solvent is also coarse-grained 3 to 1, leading to non-polar hydrophilic water beads. The lipids topology is enforced by means of 2-bodies harmonic bonded interactions and 3-bodies bending potentials. 
The zwitterionic charges are explicit. The non-bonding interactions are a combination of Lennard-Jones (LJ) and if charged, coulombic potentials. LJ interactions are cut-off at $r_c = 1.5$~nm. The coulombic potential uses a uniform static dielectric constant $\varepsilon_r = 80$. Long range electrostatics is implemented by means of the particle-particle Mesh Ewald (PPME) scheme with cut-off $r_c$. Non bonding water beads interactions have a specific form.

\subsection{Lipid compositions}

The pair of lipid compounds selected for this study is composed of di-palmitoyl-phosphatidyl-choline (DPPC) and di-linoleoyl-phosphatidyl-choline (DLiPC). DPPC  comprises a glycerol backbone with two identical saturated acyl chains of 16 carbons and phosphocholine zwitterionic headgroup. It is a cylindrically shaped lipid with an experimental gel-fluid melting temperature equal to 41$^{\circ}$C. DLiPC has the same backbone and headgroup but two double unsaturated acyl chains of 18 carbons, the \textit{cis} unsaturations being located at the 9$^{th}$ and 12$^{th}$ position ($\omega 6$, derived from linoleic acid). DLiPC is fluid at all accessible temperatures.

The disorder caused by the unsaturations is a susceptible to confer a slightly non-ideal character to this mixture while preserving a single phase mixed state. Both DPPC and DLiPC are natively included in the SPICA force-field. Our simulations were all carried out in the fluid state.

\subsection{Spica components  for the DPPC and the DLiPC molecules}


The DPPC molecule comprises 15 beads (Fig.~\ref{fig:DppcDlipcSpica}) namely NC (choline, charged +), PH (phosphate, charged -), GL (glycerol), EST1, EST2 (ester bonds), CMx (hydrophobic middle chain beads, numbered $x=1\ldots 4$ from the ester bond to the terminal methyl) and CT1, CT2 (terminal chain beads). 

The DLiPC molecule is similar in terms of bead numbers except for the 4 beads inside the acyl chains, each one covering one \textit{cis} unsaturation CMD1, CMD2.

\subsection{Simulated systems}

We prepared systems containing between 256 and 1024 lipids in total, and between 4096 and 16384 water beads. This number of lipids was chosen to comply  with the constraint of having a bilayer patch large enough to give significant results but small enough to keep  equilibration times accessible and to limit the out-of-plane roughness. These issues were mentioned in the Introduction and will be further discussed in the following section. Table~\ref{Tab:SystemCompositions} lists the systems composition used in this work. The DPPC pure system was used as a control and allowed us to define the optimal system size for the scope of this work: 512 or 1024 lipids with 8192 SPICA water molecules, which leads to approximately 24 water molecules per lipid head-group and corresponds to a satisfactory hydration state of the bilayer \cite{2021_Chattopadhyay}.
%

\begin{table}
\begin{tabular}{|c|ccc|}
\hline
Lipid &  $N_{\textrm{lipids}}$ & $N_{\textrm{wat. }}$ & Simul. time\\
\hline
DPPC &  1024/512/256 & 16384/8192/4096 & 10~$\mu$s\\
DLiPC  & 512 & 8192 & 10~$\mu$s\\
\hline
DPPC:DLiPC 3:1&  1024/512 & 8192 & 10~$\mu$s\\
DPPC:DLiPC 1:1&  1024/512 & 8192 & 10~$\mu$s\\
DPPC:DLiPC 1:3&  1024/512 & 8192 & 10~$\mu$s\\
\hline
\end{tabular}
\caption{\label{Tab:SystemCompositions}Composition of simulated systems.}
\end{table}

\subsection{Methodology}

The systems were simulated at constant temperature $T=298.15$~K using an integration time step $\delta t = 10$~fs (Lammps atomic units). After an initial preparation (using Packmol
or reusing a previous configuration) the systems were subject to the following preparation steps: energy minimization (0.5~ns), isotropic NPT barostat (50~ns, 1~atm), semi-isotropic NPT barostat (1~ns, 1~atm). After determining the optimal system box size under these pressure and temperature conditions, the production runs were performed under NVT conditions for a total of 10~$\mu$s.

The choice of using NVT rather than NPT was made in order to keep the simulations as close to the canonical ensemble from which the theory is derived as possible. As a matter of fact, we do not expect much difference between semiisotropic NPT and NVT conditions. The NVT conditions also make the computation of the density fluctuation modes (which depend on the reciprocal vectors $k_{x,y} = 2\pi (i,j)/L_{x,y}$) slightly easier, as the box size $L_x,L_y,L_z$ remains constant in time. 

Home made Python scripts (based on the MDAnalysis and mdtraj libraries) were used for analysing trajectories \textit{a posteriori} and plotting data. All visualisation was done with VMD.

\subsection{Statistical errors}

Whenever possible we tried to estimate the statistical error by computing the autocorrelation time $\tau$ of the time series that were used in the averages. This method provides a direct estimate of the 2$\sigma$ (95\%) confidence interval on a given observable, 
\begin{equation}
2\sqrt{\sigma^2}\left(\frac{\tau}{t_{\mathrm{sim}}}\right)^{1/2}
\end{equation}
where $\sigma^2$ is the variance of the time series of interest and $t_{\mathrm{sim}}$ is the simulation time. 

We also resorted to a bootstrap inductive estimator of the statistical convergence of our data. It means that our trajectories were cut \textit{a posteriori} into 10 intervals, and averages were redone by drawing at random with repetition of 100 steps new synthetic trajectories out of the 10 original fragments. The estimated variation of the resampled values was used to define the confidence interval of the observable.

\medskip
\section{The structure factor of a pure bilayer}
\label{sec:PureBilayer}
\subsection{Structure of a DPPC bilayer at the nanometer scale}

\begin{figure}
    \resizebox{0.46\textwidth}{!}{\includegraphics{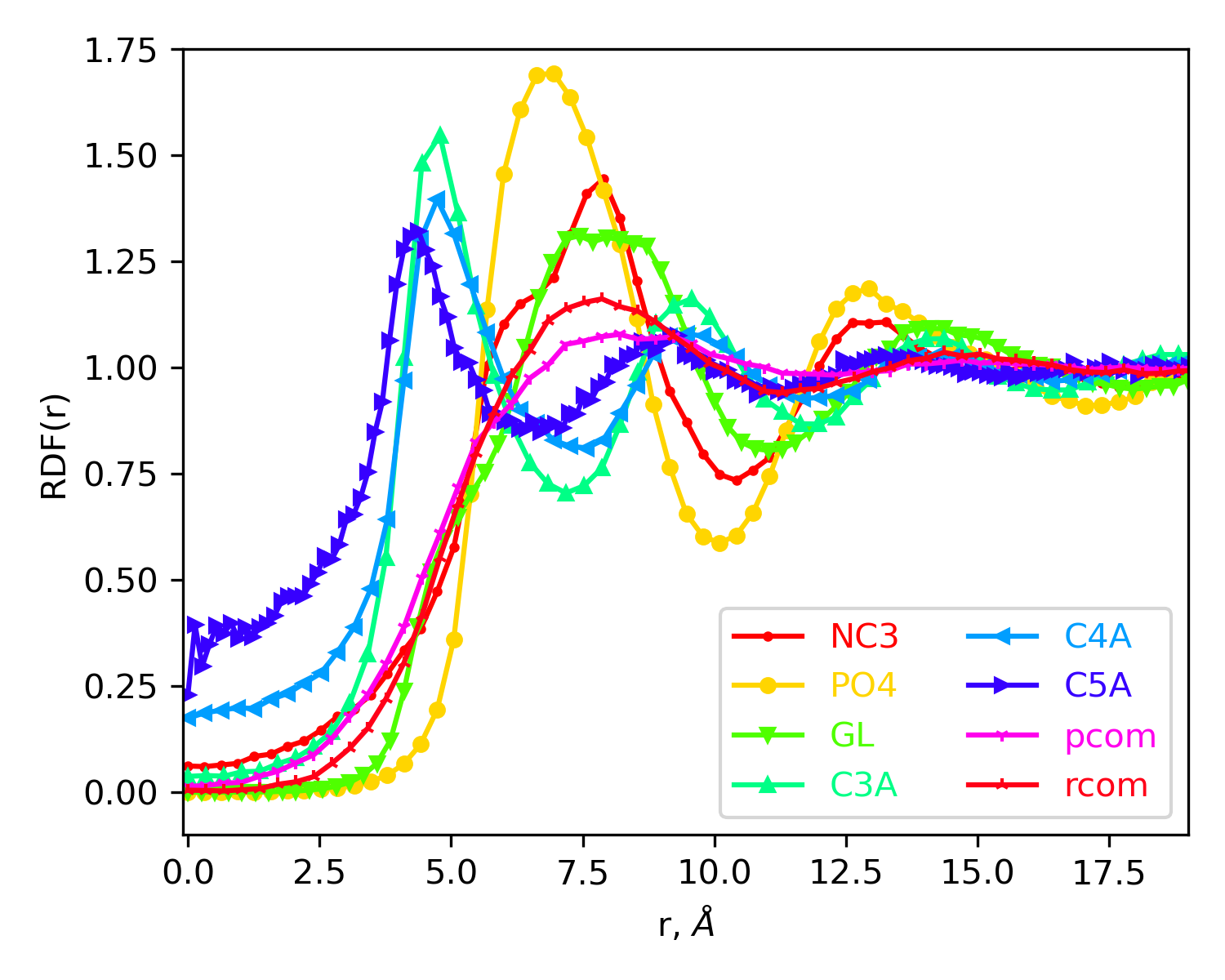}}
    \caption{\label{fig:rdfIntraLeaflet} Radial distribution function within a single leaflet of DPPC. Bead names refer to the SPICA decomposition of Fig.~\protect\ref{fig:DppcDlipcSpica}. pcom and rcom correspond to pseudo and real center of mass of the molecule as is defined in the text.}
\end{figure}

\begin{figure}
    \resizebox{0.46\textwidth}{!}{\includegraphics{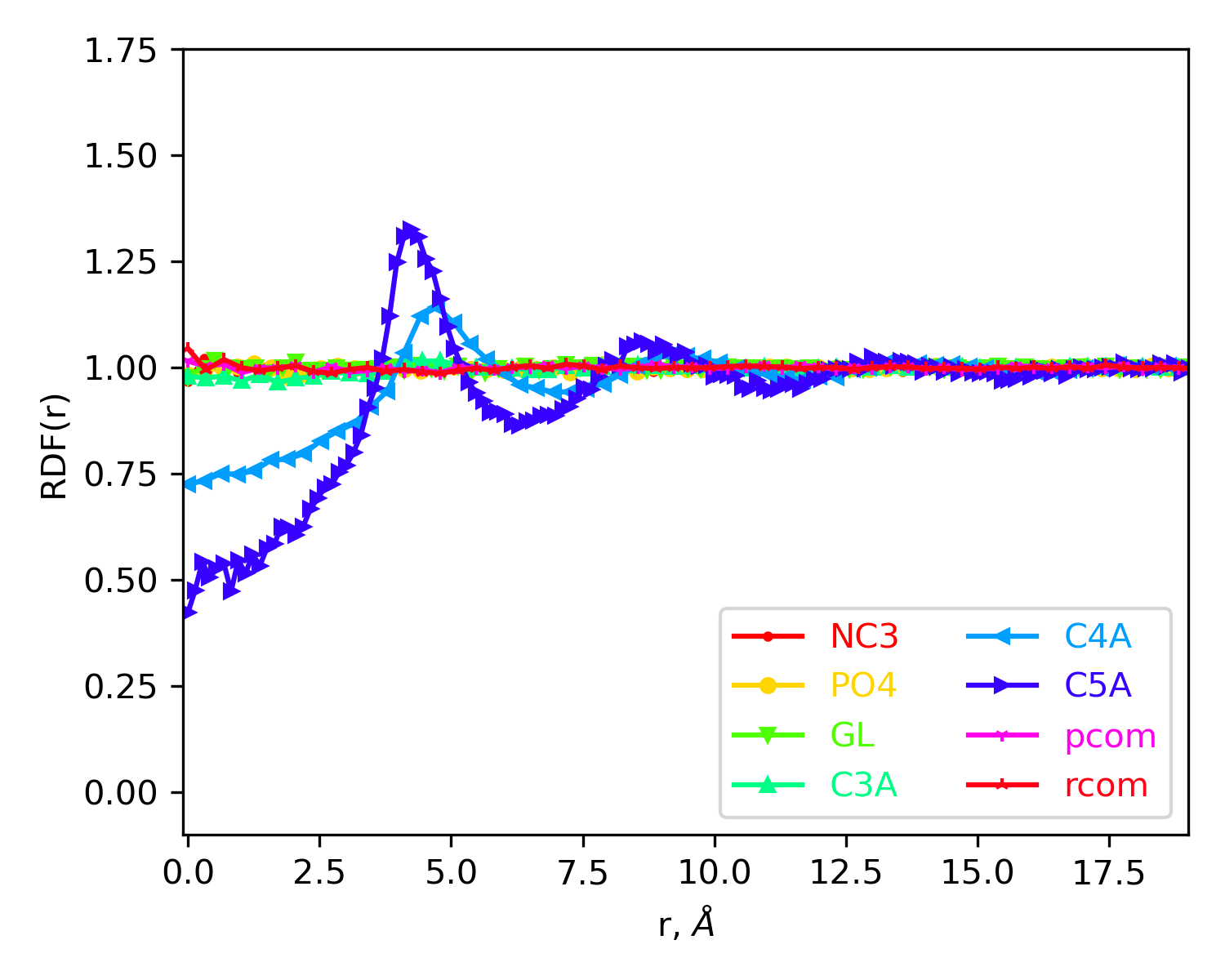}}
    \caption{\label{fig:rdfInterLeaflet} Radial distribution function between two opposite leaflets of a pure DPPC bilayer. Bead names refer to the SPICA decomposition of Fig.~\protect\ref{fig:DppcDlipcSpica}.
    }
\end{figure}


Our two-dimensional structures were obtained by projecting each lipid molecule onto a single pair of coordinates $x,y$ linked to a representative bead or a center of mass. Each frame gives rise to $N=N_l/2$ points per leaflet. These points can be used to build a radial distribution function (rdf) or pair correlation function $g(r)$. Figure~\ref{fig:rdfIntraLeaflet} shows the DPPC intra-leaflet radial distribution functions obtained for 6~different beads, and 2~center of masses. These radial distributions can be sorted into three classes. The headgroup and glycerol beads (NC3, PO4, GL) show a first density peak located around 7-7.5~\AA\; range. This is consistent with the known area per lipid of fluid DPPC bilayers.  The sharpest peak is the one of the phosphate, followed almost equally by the choline and the glycerol.
The center of mass (com) and pseudo center of mass (pcom)  offer a much less pronounced density peak but these peaks are located at the same place as the glycerol. This is due to the lack of materiality of the centers of masses and the absence of short range repulsive interactions. The effective, or mean-force, potential between lipid centers of masses is smooth and long-ranged. 
Finally, the 3 terminal beads in the tail chains show density peak at $1/\sqrt{2}$ distance from the headgroup 1st peak. This naturally reflects that lipid molecules have two identical chains. 

\begin{figure}
    \resizebox{0.46\textwidth}{!}{\includegraphics{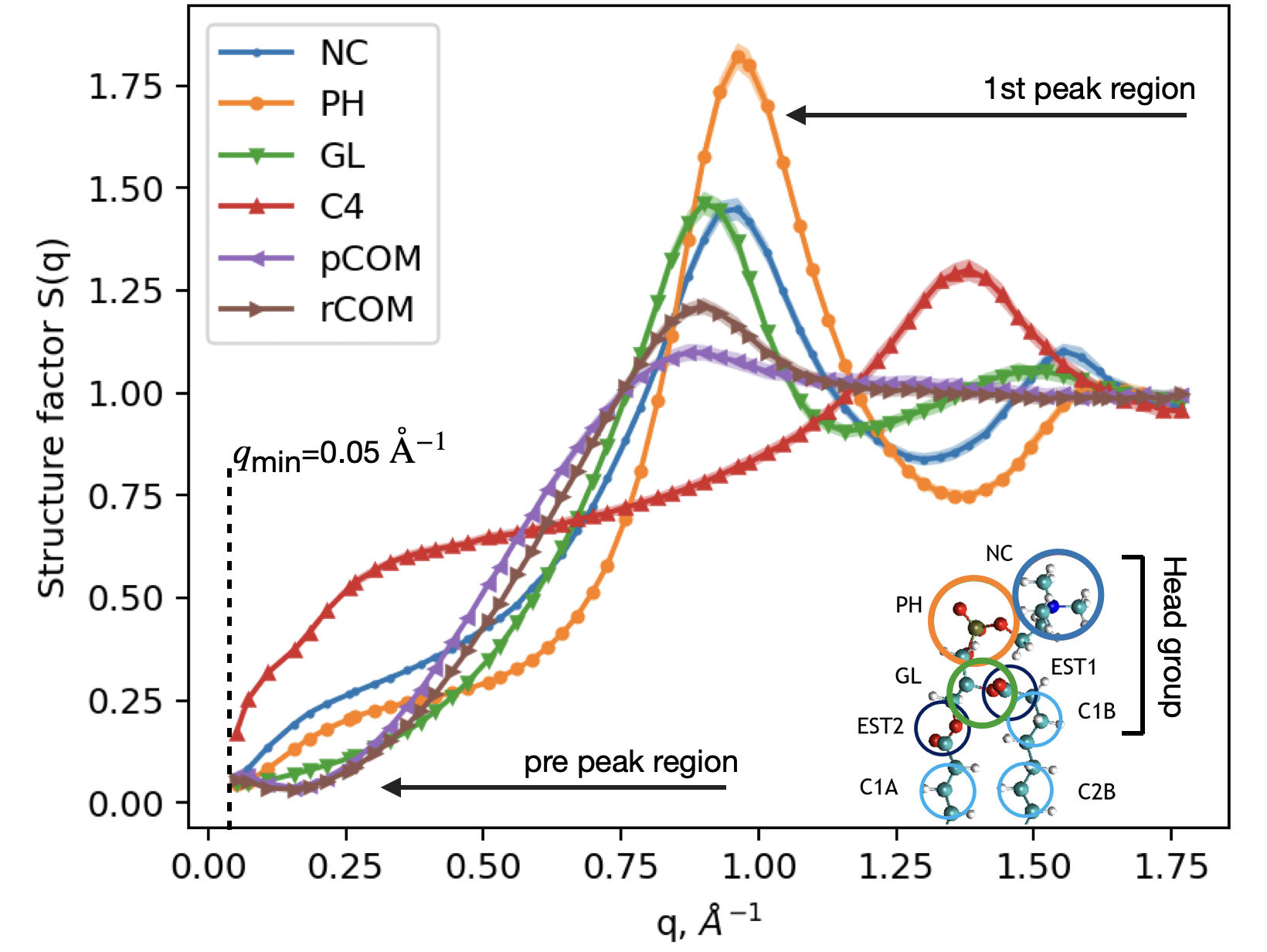}} 
    \caption{\label{fig:SqIntraLeaflet} Structure factor of a single DPPC leaflet. Bead names refer to the SPICA decomposition of Fig.~\protect\ref{fig:DppcDlipcSpica}. Head goup is shown for reference.}
\end{figure}

Figure~\ref{fig:rdfInterLeaflet} displays the correlations between molecules belonging to opposite leaflets. For all beads except the two last ones in the chains, the radial distribution is equal to~1, corresponding to an absence of positional correlation between lipids in different leaflets. Beads C4A and C5A which are located very close to the bilayer mid-plane region show a correlation peak reminiscent from the intra-leaflet radial distribution. Close to the bilayer mid-plane, the last beads in the chains tend to forget about the identity of the lipid to which they belong, and behave as a simple monomer fluid. 

Similar to the rdf, a 2d structure factor can be computed. It is shown on Figure~\ref{fig:SqIntraLeaflet}. As expected from the rdf, the three headgroup beads (PH, NC, GL) display a strong correlation peak. The sharpest peak is again the phosphate one. The centers of mass peaks are very little pronounced, and tail group beads peaks are located $\sqrt{2}$ times farther than the headgroup peaks.  

The leaflet structure factor indicates that the best locator of the lipid molecules, at this scale, is the phosphate bead. It is the one that shows the sharpest features both in real and reciprocal space. The reason why the phosphate beads are the most correlated is not obvious, but it could be a consequence of a combination of centrality (being close to the glycerol) and negative electric charge (stronger mutual repulsion). 

The shape of the structure factors at the nm$^{-1}$ scale can be rationalized quantitatively by means of a decomposition into structure and shape factors. Assuming that an optimal locator exists for each lipid molecule, and that this optimal locator is associated to an optimal sharp structure factor $S_{\mathrm{best}}(q)$, and assuming that the other beads fluctuate around the optimal locator differently depending on their positions in the molecule, then the following relation holds
\begin{equation}
S_{\mathrm{bead}}(q)-1 = (S_{\mathrm{best}}(q)-1)|F_{\mathrm{bead}}(q)|^2,
\label{eq:formFactor}
\end{equation}
with $F_{\mathrm{bead}}(q)=\exp(-i\mathbf{q} \cdot\Delta\mathbf{r})$ the term associated with the shift in position between the best locator and the actual representative bead. 
There is a close resemblance with the standard decomposition into structure and shape in the scattering analysis of a collections of identical objects, the difference (-1 term) being due to the fact that only a single bead per molecule is used for defining both $S_{\mathrm{best}}$ and $S_{\mathrm{bead}}$. The form factor of an effective intrachain harmonic spring follows a Gaussian shape, starting from~1 and decreasing to~0 on a range of the order of a few inverse molecular sizes ( 1 - 10 \AA$^{-1}$). This explains how the curves are arranged with peaks of decreasing amplitudes and moving to the left of the graph. Note that we did not attempt to fit out data to eq.~(\ref{eq:formFactor}). As we show below, the reciprocal vectors of interest for thermodynamics are not in the neighborhood of the correlation peak but rather in the low $q$ region.

\subsection{Low $q$ behavior of the bilayer structure factors}

\begin{figure}
    \resizebox{0.46\textwidth}{!}{\includegraphics{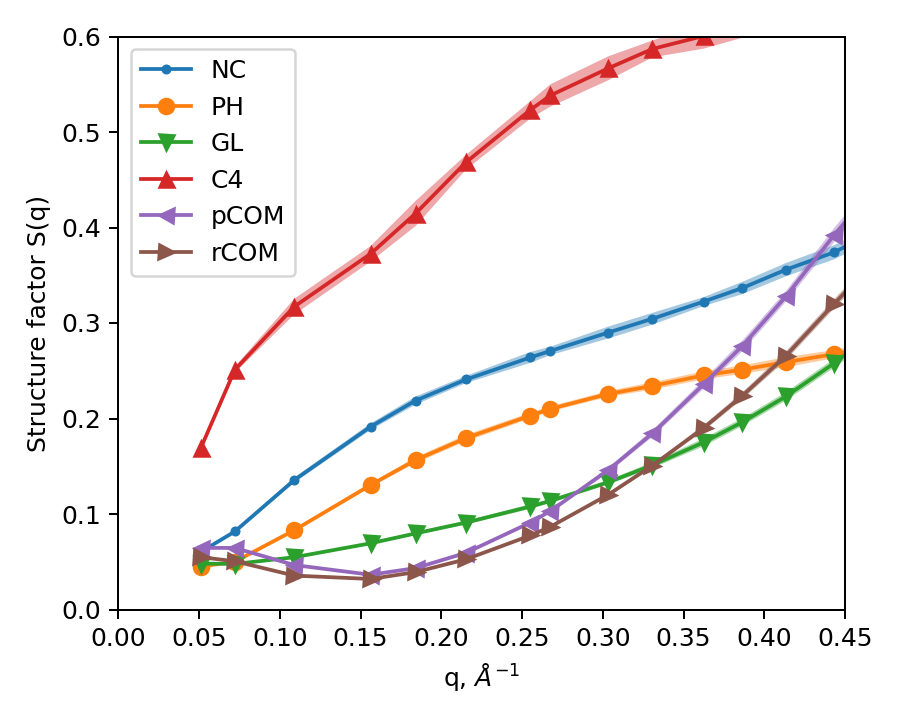}} 
    \caption{\label{fig:SqNearZero}Structure factor of a single DPPC bilayer in the hydrodynamic region. Bead names refer to the SPICA decomposition of Fig.~\protect\ref{fig:DppcDlipcSpica}.}
\end{figure}

\begin{figure}
    \resizebox{0.46\textwidth}{!}{\includegraphics{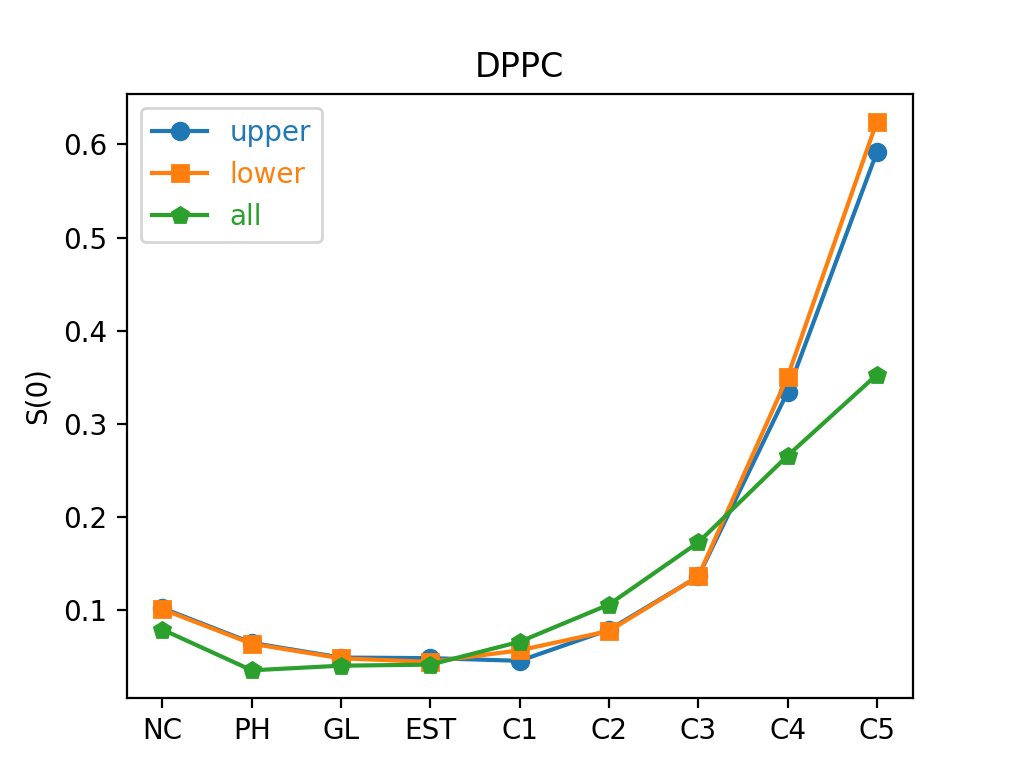}}
    \caption{\label{fig:ResolvedSq0}Extrapolated $S(0)$ values resolved in terms of SPICA beads, for the upper leaflet, the lower leaflet  and both leaflets combined. }
\end{figure}

Figure~\ref{fig:SqNearZero} represents the intraleaflet structure factors for the smallest $q$ available. There are noticeable differences between beads, NC being the more compressible, followed by PH and GL. The center of masses structure factor display a minimum and reach their $S(0)$ limit from below. Visually, all curves seem to point to a similar range of extrapolated $S(0)$ values. However, in the light of expression ~(\ref{eq:AllFluctuations}) a more systematic approach is required to capture the extrapolated $S(0)$ in each case. We therefore fit the structure factors of different beads with a quadratic $S_i(0)+C_iq^{2}$ law ($i$ being the bead index), using the $[q_{\mathrm{min}},q_{\mathrm{max}}]$ fitting range and the bootstrap estimated confidence interval for $S(q)$ associated to each $q$. The extrapolated values that we obtained are represented on Figure~\ref{fig:ResolvedSq0}. The curve is qualitatively consistent with the predictions of eq.~(\ref{eq:AllFluctuations}) with a minimum of the total fluctuations for the beads located close to the glycerol backbone (GL, EST, C1). This figure shows that both leaflets are equivalent ("upper" and "lower" on the graph). The extrapolated values associated with the combined leaflets and using eq.~(\ref{eq:BunchedReplicatedDensities}) with $\tilde{h}'=0$ corresponds to the curve "all" on the plot. The trend of the combined curve is slightly different, with a less pronounced bead dependence and a minimum shifted to the left. We attribute the flatter shape of this curve to the compensation between leaflets of the tilt and bending modes as suggested in Figure~\ref{fig:IllustrationUndulation}. The outer and inner leaflets being bent in the opposite direction, one leaflet looks expanded while the other leaflet is contracted. The combined density fluctuates therefore less than the single leaflet densities, resulting in a lower $S(q\to 0)$ value.

The bead resolved $S(0)$ values of the DLiPC molecules display similar features as the one seen for DPPC. In the subsequent part of the work, we take the GL bead as our reference for both lipids. The ester beads EST1, EST2 could have worked equally well. We also use the combination of both leaflets to mitigate the consequences of the membrane undulations. 

\subsection{Thermodynamic parameters of the pure lipid bilayers}

Eq.~(\ref{eq:compressibilityPureCase}) relates long-range density fluctuations to the bilayer area compressibility. Two other routes to determine this compressibility are 1/ a study of the box area fluctuations in a semi-isotropic barostat (area fluctuation method) and 2/ the determination of the slope of the tension-area characteristic curve obtained by imposing a non zero surface tension to the bilayer (eq. of state method). 

The box area fluctuations formula for the membrane elastic coefficient reads\cite{Heimburg_BiophysicsMembrane} 
\begin{equation}
K_A = \frac{1}{\langle A \rangle}\left( \frac{\partial \langle A\rangle}{\partial\, \sigma}\right)_{T,P} = 
\frac{k_B T\langle A\rangle }{\langle A^2\rangle - \langle A\rangle^2}.
\end{equation}
The elastic coefficient is the inverse of the compressibility coefficient and therefore
\begin{equation}
K_A = \frac{1}{\chi_T} = \frac{\rho k_B T}{S(0)}
\end{equation}
with $S(0)$ the extrapolated structure factor value. 

\begin{table}
\begin{tabular}{|c|ccc|}
    \hline
Lipid  & & $K_A^{\mathrm{str}}$ & \\
& upper$\times$2 & lower $\times 2$& bilayer  \\
    \hline
DPPC & $248.4\pm 8.4$ & $226.0\pm 11.2$ & $248.6\pm 14.4$ \\
   \hline
DLiPC & $216.6\pm 5.5$ & $215.0\pm 5.1$ & $279.8\pm 9.1$  \\
   \hline
\end{tabular}
\vspace{0.3cm}
\begin{tabular}{|c|c|c|}
    \hline
Lipid & $K_A^{\mathrm{fluct}}$ &  $K_A^{\mathrm{eqos}}$\\
    \hline
DPPC & $279.9\pm 30.6$ & $276.2\pm 14.0$ \\
\hline
DLiPC & $289.7\pm 40.0$ & $307.6 \pm 9.6$ \\
\hline
\end{tabular}
    \caption{\label{tab:StretchingElasticityPureBilayer} Values of the stretching elasticity coefficient $K_A$ for pure DPPC and DLiPC bilayers in mN/m. $K_A^{\mathrm{fluct}}$ area fluctuation method, $K_A^{\mathrm{str}}$ structure factor method, using the upper leaflet, the lower leaflet and the full bilayer, $K_A^{\mathrm{eqos}}$ equation of state method.}
\end{table}

Table~\ref{tab:StretchingElasticityPureBilayer} gives the values obtained for the membrane elastic coefficients of DPPC and DLiPC. The numerical values are consistent with the experimental values ($231\pm 20$~mN/m for DPPC, $247 \pm 21$~mN/m for DLiPC~\cite{Marsh_HandbookLipidBilayers2, 2000_Rawicz_Evans}). 

The structure factor values derived from single leaflets are smaller than the one derived from the full bilayer. This means that the extrapolated $S(0)$ values are larger for single leaflets than for the full bilayer. We interpret this as the consequence of an incomplete subtraction of the lipid tilt and undulation contributions in the single leaflets case. The value obtained for the combined leaflets (full bilayer) is closer to the box area fluctuation method. The box fluctuation method has large error bars, and is lower but consistent with the equation of state method. We conclude that the structure factor approach tends to slightly underestimate the elastic coefficient parameter, which is likely due to an incomplete removal of the tilt and undulation contributions. The agreement remains satisfactory.  


\medskip
\section{Binary lipid mixtures}
\label{sec:Mixture}

\subsection{Issues with simulations of lipid mixtures}

The adaptation of the density modes fluctuations approach to binary mixtures comes with a few specific difficulties. The study of the mixing properties requires 3 mutual density modes $\tilde{H}_{11},\tilde{H}_{12}$ and $\tilde{H}_{22}$. The time needed to gather the necessary statistics increases a lot as one of the lipid species becomes a minor component, either $x\to 0$ or $x\to 1$. In practice, one is restricted to simulate mixtures that do not depart to much from the equimolar proportion. In this work we restricted ourselves to 3:1, 1:1 and 1:3 lipid ratios.

A second major difficulty arises from the fact that the intrinsic correlation times of the $\hat{n}_{\mathbf{q},\alpha}(t)$ modes is much slower than in the pure system case. In the mixture, the thermalization of the density fluctuations occurs through a lengthy self-diffusion process where molecules of one species must overcome the molecular friction of the adverse species. In the hydrodynamic regime, the thermodynamic forces (gradient of chemical potentials) are weak and this friction is high. By contrast, in the pure system, the density fluctuations are controlled by a collective diffusion coefficient for which the relative molecular motions plays no part. Apart from a moderate internal friction, only interactions with the solvent or the opposite leaflet can slow down the collective fluctuation dynamics. 

\subsection{Using fake mixtures as a benchmark}
As we have learned from pure mixtures and before turning to real mixtures, there is an intermediate situation of interest, where one takes a pure system and relabel a posteriori the lipids as if they belonged to two different species. The binary mixture obtained in this way is by construction ideal, because the molecules are perfectly equivalent and substituable. Yet this is not a trivial system. We refer to such systems as \textit{fake mixtures}.

The formalism should yield a consistent vanishing $B$ coefficient irrespective of the "lipid composition". In what follows, we compare the $B$ obtained for the true DPPC-DLiPC mixture and the ones obtained for a fake DPPC mixture. 

\begin{figure}
    \resizebox{0.46\textwidth}{!}{\includegraphics{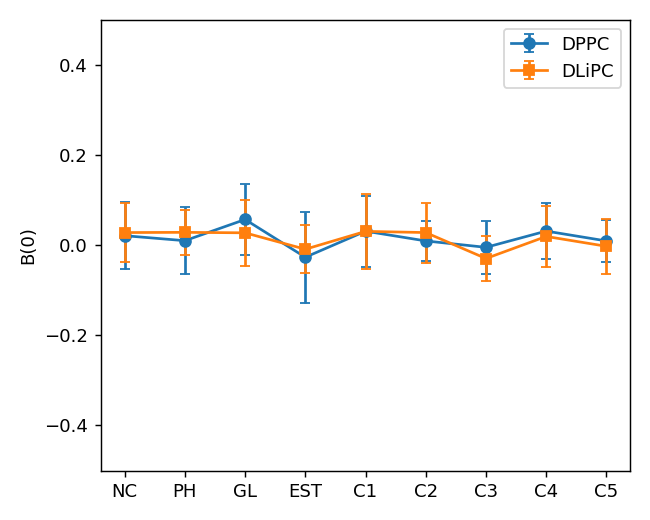}} 
    \caption{\label{fig:FakeMixtureDPPC}Non-ideal mixing parameter $B$ for a 50:50 fake DPPC and DLiPC mixtures, taking different beads as lipid reference positions. }
\end{figure}

Figure~\ref{fig:FakeMixtureDPPC} shows that the non-ideal mixing parameter is much less sensitive to the choice of the bead than the structure factor $S(0)$. The resulting $B$ values are close to 0, and $B=0$ is consistent with our estimated error bars. It is likely that the collective tilt and undulation contributions cancel out in the parameter $\Delta$ appearing in eq.~(\ref{eq:KBRelationsBinaryMixtures}). 

\subsection{Non-ideality of DPPC-DLiPC bilayers}

We are now in position to compute the non-ideal mixing parameter of a numerical mixture of DPPC and DLiPC. The simulated systems comprises 1024 lipids, and the full bilayer statistics is used to evaluate $B$. 

\begin{figure}
    \resizebox{0.46\textwidth}{!}{\includegraphics{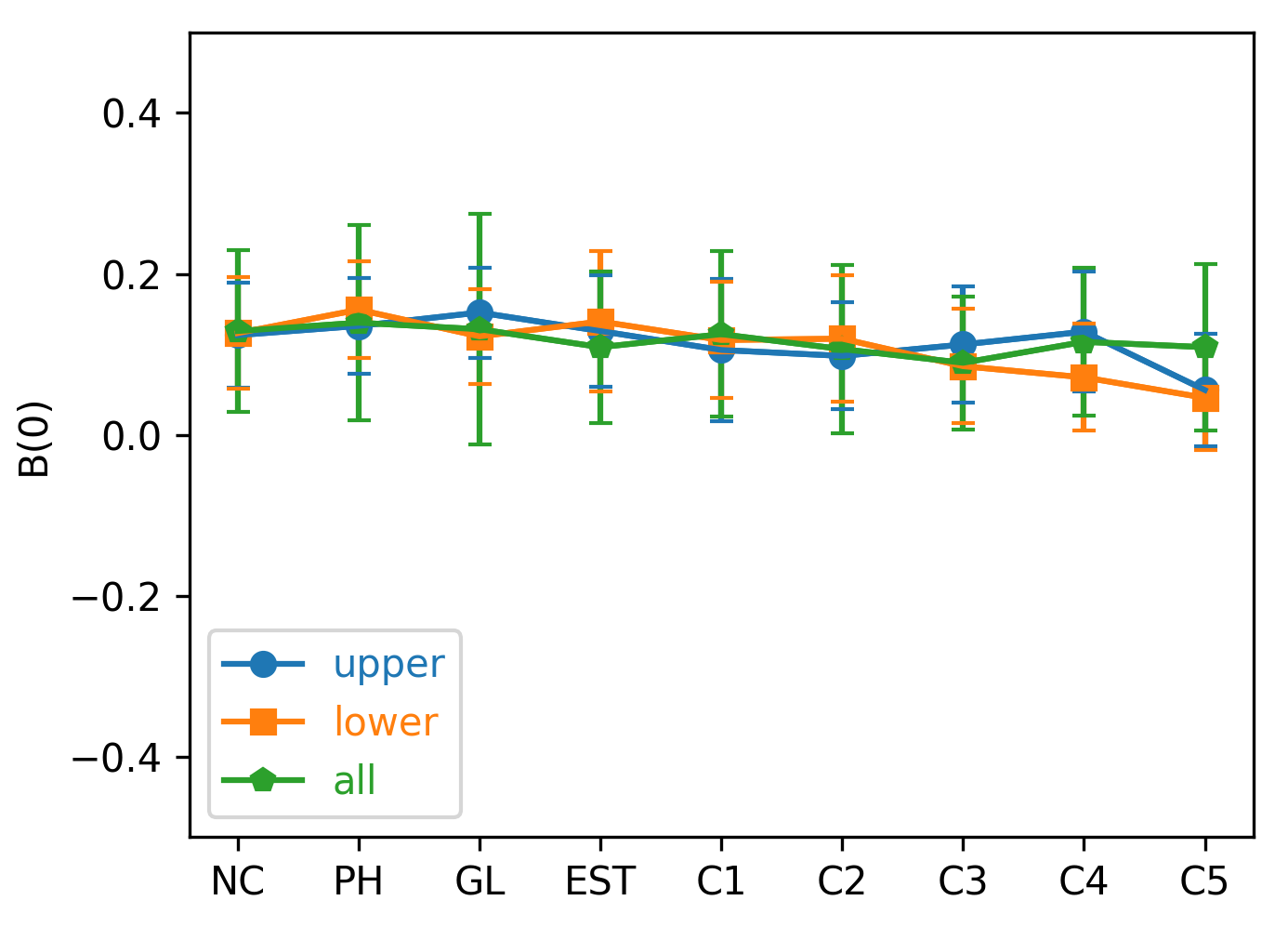}}
    \caption{\label{fig:DPPCDLiPC5050} Non-ideal mixing parameter $B$ for a 50:50 DPPC-DLiPC fluid mixture, taking different beads as lipid reference positions, using either single leaflets  or both leaflets combined. }
\end{figure}

Figure~\ref{fig:DPPCDLiPC5050}
shows the non-ideal mixing parameter $B$ obtained in an equimolar mixture of DPPC and DLiPC. This parameter is again relatively insensitive to the choice of the reference bead. Our numerical estimate for $B$ is close to 0.1, which represents a weak tendency to demix, though quite far away from the critical separation value. DPPC and DLiPC mix really well, in a  slightly non-ideal way. 

\begin{figure}
    \resizebox{0.46\textwidth}{!}{\includegraphics{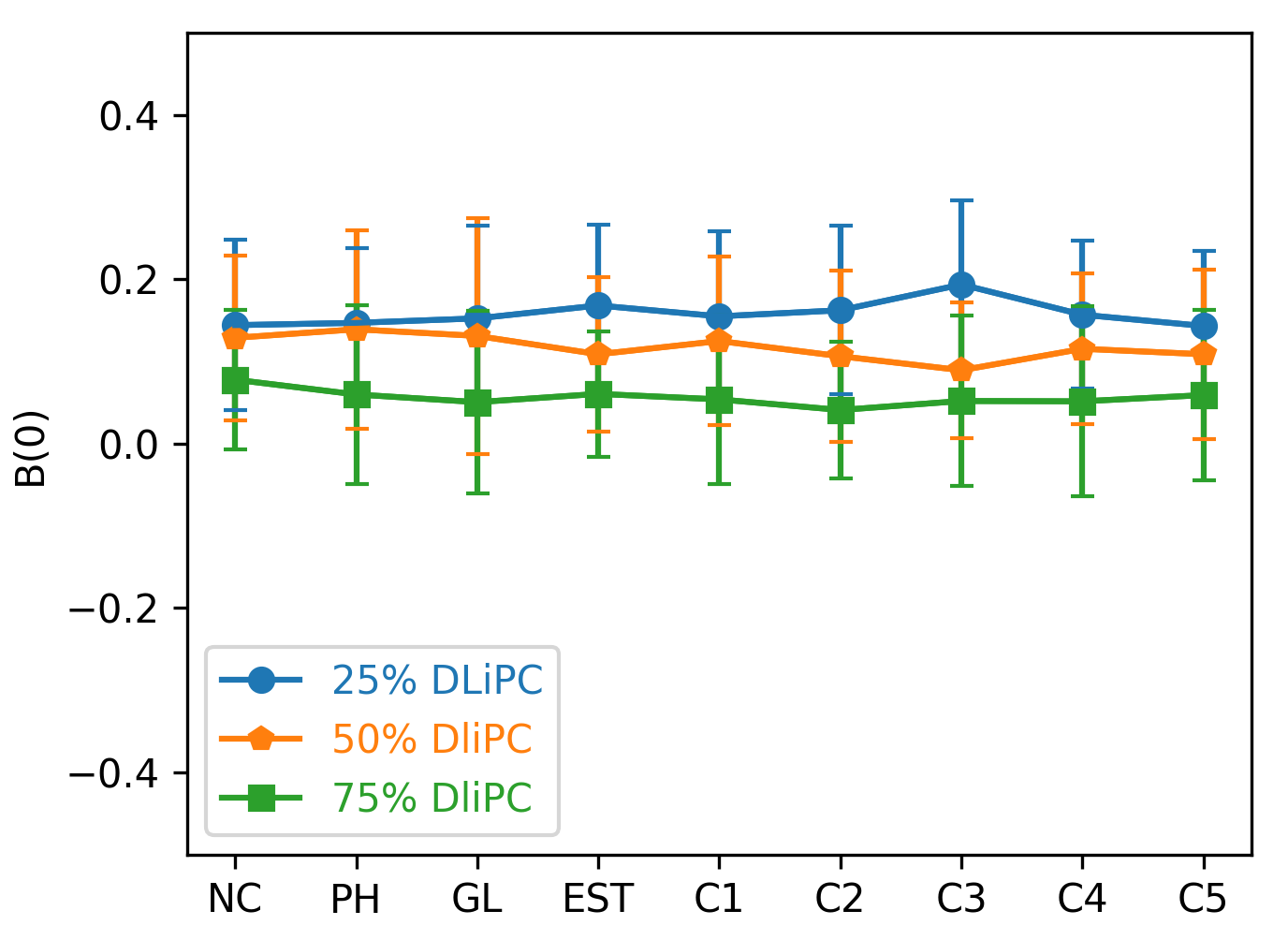}}
    \caption{\label{fig:DPPCDLiPC}Non-ideal mixing parameter $B$ for 3 compositions of a DPPC-DLiPC fluid mixture: 25:75, 50:50 and 75:25.}
\end{figure}

\begin{figure}
   \resizebox{0.46\textwidth}{!}{\includegraphics{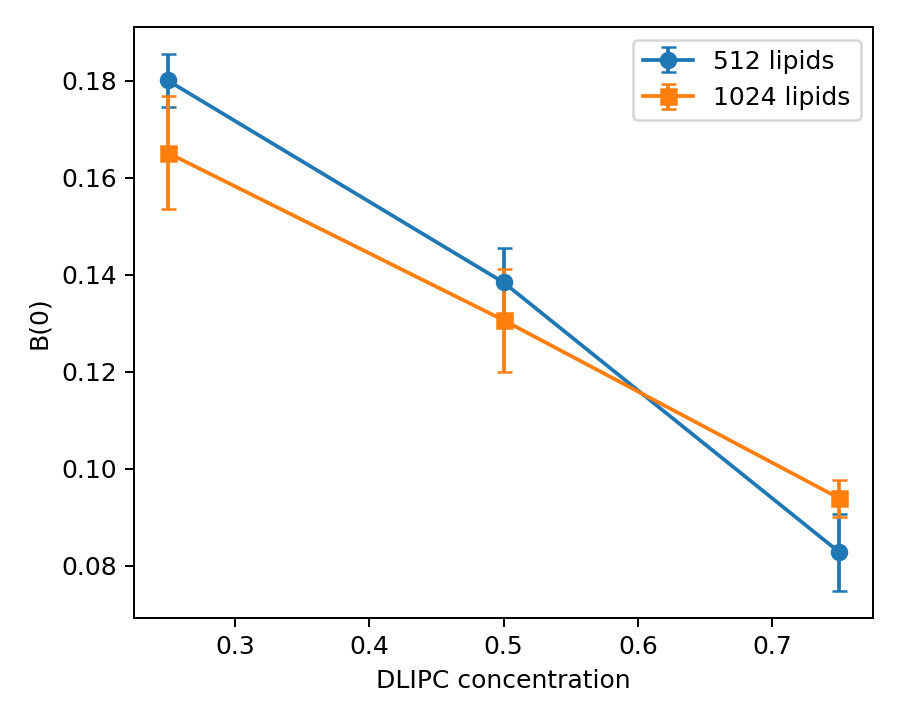}}
   \caption{\label{fig:MixingEvolution}Evolution of the non-ideal mixing parameter with the mixture composition for two studied system sizes. Averaging done over GL and EST beads over upper, lower leaflets and the whole bilayer.}
\end{figure}

Figure~\ref{fig:DPPCDLiPC} superimposes the non-ideal mixing parameter obtained for the 3 studied compositions ($x=0.25,~0.5,~0.75$). Again, the parameter $B$ is rather well defined for a given composition. Its numerical value decreases as the concentration of DLiPC increases, as represented on Figure~\ref{fig:MixingEvolution}. The observed decrease in the value of $B$ with $x$ indicates a deviation from the quadratic theory of regular solutions. $B(x)$ remains however consistently and significantly positive. 

\subsection{Consistency of the thermodynamic values}

\begin{table}
\begin{tabular}{|c|ccc|}
    \hline
Lipid mixture  & & $K_A^{\mathrm{str}}$ & \\
& upper$\times$2 & lower $\times 2$& bilayer  \\
    \hline
DPPC/DPPC & $253.5\pm 7.0$ & $256.6\pm 6.7$ & $275.8\pm 10.3$ \\
DLiPC/DLiPC & $210.6\pm 15.5$ & $220.4\pm 14.9$ & $278.6\pm 18.0$  \\
   \hline
DLiPC/DPPC 1:3 & $242.7\pm 4.2$ & $242.9\pm 4.1$ & $293.1\pm 7.4$ \\
DLiPC/DPPC 1:1 & $232.6\pm 2.5$ & $231.0\pm 9.8$ & $280.8\pm 9.1$ \\
DLiPC/DPPC 3:1 & $222.0\pm 3.9$ & $224.6\pm 3.0$ & $279.3\pm 9.0$ \\
\hline
\end{tabular}

\vspace{0.3cm}
\begin{tabular}{|c|c|c|}
    \hline
Lipid mixture & $K_A^{\mathrm{fluct}}$ &  $K_A^{\mathrm{eqos}}$\\
    \hline
DPPC/DPPC & $279.9\pm 30.6$ & $276.2\pm 14.0$ \\
DLiPC/DLiPC & $304.6\pm 31.5$ & $314.2 \pm 12.8$ \\
    \hline
DLiPC/DPPC 1:3 & $270.6\pm 19.6$ & $291.7\pm 10.1$\\
DLiPC/DPPC 1:1 & $288.4 \pm 35.4$ & $303.0 \pm 9.8$ \\
DLiPC/DPPC 3:1 & $289.7\pm 40.0$ & $307.6 \pm 9.6$  \\  
   \hline
\end{tabular}
    \caption{\label{tab:StretchingElasticityMixtures}Values of the stretching elasticity coefficient $K_A$ for mixtures of DPPC and DLiPC in mN/m. $K_A^{\mathrm{fluct}}$ area fluctuation method, $K_A^{\mathrm{str}}$ structure factor method, using the upper leaflet, the lower leaflet and the full bilayer, $K_A^{\mathrm{eqos}}$ equation of state method. The two first lines are fake mixtures of DPPC and DLiPC,  $K_A^{\mathrm{fluct}}$ and $K_A^{\mathrm{eqos}}$  are therefore equal to the values quoted in Table~\protect\ref{tab:StretchingElasticityPureBilayer}.}
\end{table}

We finally checked whether the thermodynamic parameters of the bilayer obtained using eq.~(\ref{eq:KBRelationsBinaryMixtures}) were consistent. 
Table~\ref{tab:StretchingElasticityMixtures} presents the stretching elastic coefficients of the 2~fake mixtures and the 3 real mixtures. 
The values obtained for the 2~fake mixtures are consistent with the values in Table~\ref{tab:StretchingElasticityPureBilayer}. The error bars of $K_A^{\mathrm{str}}$ are larger than in the pure case, reflecting the decrease in the number of lipids of each kind. The compressibility values 
$K_A^{\mathrm{str}}$ derived from the combined  leaflets are all closer to the other methods, and larger than the values corresponding to single leaflets. The fluctuations at $q\to 0$ are more important for the single leaflets than for the combined leaflets. The area fluctuation method suffers from large error bars, due to the large fluctuations of the variance of the area (variance of the variance). The equation of state values always lie above the structural values by \textit{ca} 10\%. We again attribute it the to incomplete substraction of the tilt and undulations contributions. Altogether, the compressibility (or elastic coefficient) of the mixtures agrees well with the pure case ones. We therefore conclude that our data supports the use of  eq.~(\ref{eq:KBRelationsBinaryMixtures}).

\medskip
\section{Discussion, Methodology, Perspectives}
\label{sec:Methodology}

In the previous section, we were able to obtain a quantitative estimate of the quadratic thermodynamic of mixing parameter from MD simulations of homogeneous binary lipid mixtures. The order of magnitude of this parameter in the case under study was found to be low ($\sim 0.1$) and far from the region where critical fluctuations or demixing are expected to occur. This is not surprising given the similarity between the two simulated species, which are known to mix well experimentally. Our coarse-grained approach is in addition expected to smoothen molecular details and may contribute to increase the similarity between lipid molecules as compared with their atomistic representations. 
We also observed that the quadratic mixing parameter depends on the concentration, possibly pointing to a deviation from simple quadratic regular solutions. 

The main difficulty to overcome in practice is the large size of the simulation boxes necessary to extrapolate the density fluctuations to their long wave $q \to 0$ limit. Simulating large boxes requires a bigger computational effort, with more computing nodes and longer times needed to thermalize the slow hydrodynamic modes. We believe that we successfully found an optimal compromise by simulating systems comprising between 512 and 1024 lipids. In these systems we could obtain a consistent trend for the extrapolated $\tilde{h}_{\alpha\beta}(q \to 0)$ values of the equilibrium density modes fluctuations, as far as the lipid representative beads were concerned. We are reasonably confident that the use of the central glycerol bead (GL) allowed us to subtract unwanted contributions of the membrane undulations.

It is common to explain and rationalize the non-ideal interactions between molecules by reasoning with a lattice-gas model and nearest neighbor coupling parameters. A 2d binary mixture can for instance be mapped to particles occupying the vertices of an hexagonal lattice. Within this picture, each site is surrounded by $z=6$ nearest neighbors. A bond connecting two nearest neighbors contributes by an amount $\omega_{11}$, $\omega_{22}$ or $\omega_{12}$ to the total free-energy of the system. The energies $\omega_{11}$, $\omega_{22}$ and $\omega_{12}$ correspond respectively to a bond separating two lipids of type~1, two lipids of type~2 or a couple of different lipids. The analysis of binary lattice gases shows that the parameter controlling the thermodynamics of mixing is the combination $\omega = \omega_{12}-(\omega_{11}+\omega_{22})/2$. Positive $\omega$ values are associated with unfavorable mixing. 

Connection with our approach is possible far from the critical value of demixing       by means of the correspondence 
\begin{equation}
B = \frac{z\omega}{RT} = \frac{6\omega}{RT}
\end{equation}
with $z=6$ corresponding to an hexagonal lattice and $\omega$ expressed in J.mol$^{-1}$. Therefore $B\simeq 0.1$ can be interpreted as an unfavorable nearest neighbour mixing of $\omega\simeq 41~\mathrm{J.mol}^{-1}\simeq 10~\mathrm{cal.mol}^{-1}$. Such values can be compared with values inferred from experiments or Monte-Carlo (MC) simulations aiming at reproducing experimental features. However, we could not find experimental estimates for the DPPC-DLiPC binary mixture in the data compiled by Almeida~\cite{2009_Almeida}. 


Preliminary results indicate that values of $B$ of order unity can be observed between DPPC and glycolipid GCER (Martini model introduced in~\cite{2013_Lopez_Marrink}). Larger $B$ values are indeed easier to measure thanks to a larger signal to noise ratio in the relative density fluctuations. 
Similarly the interaction of saturated lipid DPPC with cholesterol gives rise to strong negative values of $B$. This is expected by virtue of the so-called umbrella effect, a well established effect that tends to appariate cholesterol to saturated lipid components \cite{1999_Huang_Feigenson_2, 1999_Radhakrishnan_McConnell}. While a theory of regular solutions may not be appropriate to describe cholesterol lipid interactions, the emergence of correlated concentration fluctuations between the two species is clearly physical and leads indeed to negative $B$ values.

The results presented in this manuscript allows us to outline a general procedure to extract non-ideality parameters from coarse-grained molecular dynamics simulations. The steps are as follows.
\begin{enumerate}
\item Simulate first pure bilayers of each lipid species. The system size should comprise 512-1024 lipids.
\item Compute the density fluctuation spectrum of each representative bead, with both separated and combined leaflets. Extrapolate to $q\to 0$ the fluctuations as in Figure~\ref{fig:ResolvedSq0}.
\item Select the beads which minimize the extrapolated density fluctuations 
\item Simulate binary mixtures of lipids with symmetric leaflet composition. Obtain~$B$ and check for dependence in composition and bead choice.
\item Perform a control simulation by randomly relabeling the pure systems (fake mixtures).
\end{enumerate}

Further work is required to fill the full matrix of lipid pairs interactions, and to extend the approach to atomistic lipid models. This should be doable provided one have access to computing ressource large enough to simulate systems equivalent to the ones used in this work.


\medskip
\section{Conclusion}
\label{sec:Conclusion}

We presented a procedure to compute the non-ideal thermodynamic of mixing parameter of a binary lipid mixture. This method relies on the analogy between a lipid bilayer and two nearly independent coupled 2-dimensional fluids. From the statistical thermodynamics of binary fluids we derive an expression for the derivative of the chemical potential of one species with respect to the other based on density modes fluctuations analysis. 
Under the assumption that the mixture is well described by a theory of regular solution, the chemical potential derivative leads to a quantitative expression for the non-ideal mixing parameter. Applied to the SPICA model of DPPC-DLiPC, we obtained a value of $B$ of the order of $0.1$, indicating nearly ideal mixing.

Mapping a lipid bilayer to a flat 2d fluid can be considered as an extreme coarse-graining step and a significant simplification of reality. In addition, membrane undulations make it difficult to properly extrapolate the longitudinal density fluctuation values to the thermodynamic limit. A better description would involve the theoretical treatment of binary fluids of particles embedded in a fluctuation 2d manifold, a quite challenging perspective. 

Our approach is also currently restricted to bilayers of symmetric compositions. In order to extend the approach to asymmetric membranes, it will be necessary to consider quaternary mixtures and deal properly with the tilt-composition couplings. This seems a reasonable endeavour.


\section*{Acknowledgements}

R.~Kociurzynski acknowledges support from the Graduate school IRTG Soft Matter Science (SoMaS). This work was performed using HPC resources from GENCI–IDRIS (Grant 2022-A0120712495).

\medskip
\appendix
\section{Theoretical complements}
\label{sec:Appendix}

\subsection{Expressions of the compressibility $\chi_T$ and the mixing parameter~$B$}

We consider a flat 2d binary mixture of composition $N_1$, $N_2$, area $A$ and temperature $T$. The system thermodynamics is characterized by a Helmholtz free-energy state function $F(N_1,N_2,A,T)$ and we assume that the fluctuation relations~(\ref{eq:GrandCanonicalNumberOfParticles}) and~(\ref{eq:GrandCanonicalKBPure}) hold. 
In all the following calculations \textit{the temperature~$T$ is kept constant}, and omitted from the notations relative to partial derivatives.

Introducing the densities $\rho_1 = N_1/A$, $\rho_2=N_2/A$ and $f = F/A$, the homogeneity of the free-energy $F$ leads to the relation 
\begin{equation}
F(N_1,N_2,A,T) = A f(\rho_1,\rho_2,T),
\end{equation}
from which one deduces the two chemical potentials 
\begin{equation}
\mu_1 = \left(\frac{\partial f}{\partial \rho_1}\right)_{\rho_2}\; ; \; 
\mu_2 = \left(\frac{\partial f}{\partial \rho_2}\right)_{\rho_1}, 
\end{equation}
and the system tension $\sigma$ (opposite of a pressure)
\begin{eqnarray} 
    \sigma &=&\left(\frac{\partial F}{\partial A}\right)_{N_1,N_2} \nonumber\\
    &=&  \left(\frac{\partial Af(N_1/A,N_2/A,T)}{\partial A}\right)_{N_1,N_2}\nonumber\\
    &=& f-\rho_1\frac{\partial f}{\partial \rho_1}-\rho_2\frac{\partial f}{\partial \rho_2}.
    \label{eq:LegendreSigmaf}
\end{eqnarray}
The tension $\sigma$ and the free-energy density $f$ are therefore related by a Legendre transform. This relation also expresses the relation between the Gibbs ($G=\mu_1 N_1+\mu_2N_2$) and the Helmholtz ($F=G+\sigma A$) free-energies. 

The tension $\sigma$ is connected to the system grand potential $\mathcal{J}(\mu_1,\mu_2,A,T)$ by the relation $A\sigma(\mu_1,\mu_2,T) = \mathcal{J}$. The densities $\rho_1$ and $\rho_2$ can thus be obtained from $\mathcal{J}$ or $\sigma$:
\begin{eqnarray} 
N_1 &=& -\left(\frac{\partial \mathcal{J}}{\partial \mu_1}\right)_{\mu_2,A}\nonumber\\
\rho_1 &=& -\left(\frac{\partial \sigma}{\partial \mu_1}\right)_{\mu_2}
\end{eqnarray}
and similarly 
\begin{equation}
\rho_2 = -\left(\frac{\partial \sigma}{\partial \mu_2}\right)_{\mu_1}
\end{equation}
The $2\times 2$ matrices 
\begin{equation}
\left(\frac{\partial \mu_{\alpha}}{\partial \rho_{\beta}}\right)=\left(\frac{\partial^2 f}{\partial \rho_{\alpha}\partial \rho_{\beta}}\right)
\end{equation}
and 
\begin{equation}
\left(\frac{\partial \rho_{\alpha}}{\partial \mu_{\beta}}\right)=-\left(\frac{\partial^2 \sigma}{\partial \mu_{\alpha}\partial \mu_{\beta}}\right)
\end{equation}
are matrix inverses. Eq.~(\ref{eq:GrandCanonicalKBPure}) establishes precisely a link between the KB integrals and $-\partial^2 \sigma/\partial \mu_{\alpha}\partial \mu_{\beta}$:
\begin{equation}
-\left(\frac{\partial^2 \sigma}{\partial \mu_{\alpha}\partial \mu_{\beta}}\right) 
 = \frac{1}{k_b T}\left(%
 \begin{array}{cc}
 \rho_1 + \rho_1^2 G_{11} & \rho_1\rho_2 G_{12}\\
 \rho_1\rho_2 G_{12} & \rho_2 + \rho_2^2 G_{22}
 \end{array}
 \right)%
\end{equation}
from which we obtain
\begin{equation}
\left(\frac{\partial^2 f}{\partial \rho_{\alpha} \partial \rho_{\beta}}\right) 
 = \frac{k_b T}{\rho_1\rho_2 \zeta} \left(
 \begin{array}{cc}
 \rho_2 + \rho_2^2 G_{22} & -\rho_1\rho_2 G_{12}\\
 -\rho_1\rho_2 G_{12} & \rho_1 + \rho_1^2 G_{11}
 \end{array}
 \right),
 \label{eq:secondDerivativesf}
\end{equation}
with $\zeta$  defined as 
\begin{equation}
\zeta = 1+\rho_1 G_{11}+\rho_2 G_{22}+\rho_1\rho_2(G_{11}G_{22}-G_{12}^2).
\label{eq:definitionZeta}
\end{equation}
To obtain the compressibility, one writes
\begin{eqnarray}
\frac{1}{A\chi_T} &=& \left(\frac{\partial \sigma}{\partial A}\right)_{N_1,N_2} \nonumber\\
&=& \left(\frac{\partial\left(f-\rho_1\frac{\partial f}{\partial \rho_1}-\rho_2\frac{\partial f}{\partial \rho_2}\right)}{\partial \rho_1}\right)_{\rho_2} \times \left(\frac{\partial \rho_1}{\partial A}\right)_{N_1,N_2}\nonumber\\
& & + \left(\frac{\partial \left(f-\rho_1\frac{\partial f}{\partial \rho_1}-\rho_2\frac{\partial f}{\partial \rho_2}\right)}{\partial \rho_2}\right)_{\rho_1} \times \left(\frac{\partial \rho_2}{\partial A}\right)_{N_1,N_2}\nonumber\\
\frac{1}{\chi_T}&=& \rho_1^2 \frac{\partial^2 f}{\partial \rho_1^2}+\rho_2^2 \frac{\partial^2 f}{\partial \rho_2^2}+
2\rho_1\rho_2 \frac{\partial^2 f}{\partial \rho_1\rho_2}
\label{eq:compressibilityBinaryMixture}
\end{eqnarray}
Combining (\ref{eq:compressibilityBinaryMixture}), (\ref{eq:definitionZeta}) and (\ref{eq:secondDerivativesf}) leads to an expression for the inverse compressibility
\begin{equation}
\frac{1}{\chi_T} = \frac{k_B T \eta}{\zeta}
\end{equation}
with $\eta$ defined as:
\begin{equation}
\eta = \rho_1+\rho_2 + \rho_1\rho_2(G_{11}+G_{22} - 2G_{12})
\end{equation}

The matrix (\ref{eq:secondDerivativesf}) provides a straightforward expression for the derivative of the chemical potential at constant area $(\partial \mu_1/\partial N_2)_{A,T}$. Unfortunately as expression~(\ref{eq:MuCrossDerivative}) shows, what is required when working with a regular solution description is the derivative of the chemical potential at constant tension ~$\sigma$. To express the latter in terms of the former, it is convenient to first define the specific area
\begin{equation}
\mathcal{A}_{\alpha}= \left(\frac{\partial A}{\partial{N_{\alpha}}}\right)_{N_{\alpha}',\sigma}
\end{equation}
using the relation between the 3 implicitly dependent variables $A,\sigma,N_{\alpha}$
\begin{equation}
\left(\frac{\partial A}{\partial N_{\alpha}}\right)_{\sigma}\times \left(\frac{\partial \sigma}{\partial A}\right)_{N_{\alpha}}\times\left(\frac{\partial N_{\alpha}}{\partial \sigma}\right)_{A} = -1
\label{eq:intermediaire1}
\end{equation}
with 
\begin{equation}
\left(\frac{\partial \sigma}{\partial A}\right)_{N_{\alpha}} = \frac{1}{A\chi_T}
\label{eq:intermediaire2}
\end{equation}
and 
\begin{eqnarray}
\left(\frac{\partial \sigma}{\partial N_{\alpha}}\right)_{N_{\alpha}} &=& \frac{1}{A} \left(\frac{\partial\sigma}{\partial \rho_{\alpha}}\right)_{\rho_{\alpha}'}\nonumber\\
&=& \frac{1}{A} \left(\frac{\partial \left(f-\rho_1\frac{\partial f}{\partial \rho_1}-\rho_2\frac{\partial f}{\partial \rho_2}\right)}{\partial \rho_{\alpha}}\right)_{\rho_{\alpha}'}
\nonumber\\
&=& -\frac{1}{A}\sum_{\beta} \rho_{\beta} \frac{\partial^2 f}{\partial \rho_{\alpha}\partial \rho_{\beta}}.
\end{eqnarray}
Leading to
\begin{eqnarray}
\left(\frac{\partial \sigma}{\partial N_1}\right)_{N_{\alpha},T} &=& -\frac{1}{A} \left(\rho_1 \frac{\partial^2 f}{\partial \rho_1^2} + \rho_2   \frac{\partial^2 f}{\partial \rho_1\partial\rho_2}\right)\nonumber\\
\left(\frac{\partial \sigma}{\partial N_2}\right)_{N_{\alpha},T} &=& -\frac{1}{A} \left(\rho_2 \frac{\partial^2 f}{\partial \rho_2^2} + \rho_1   \frac{\partial^2 f}{\partial \rho_1\partial\rho_2}\right)
\end{eqnarray}
and finally 
\begin{eqnarray}
\mathcal{A}_1 &=& \frac{1+\rho_2(G_{22}-G_{12})}{\rho_1+\rho_2 + \rho_1\rho_2(G_{11}+G_{22}-2G_{12})}\nonumber\\
\mathcal{A}_2 &=& \frac{1+\rho_1(G_{11}-G_{12})}{\rho_1+\rho_2 + \rho_1\rho_2(G_{11}+G_{22}-2G_{12})}
\end{eqnarray}

The connection between derivative of the chemical potential at constant area and constant tension comes from the differentiation of the identity
\begin{equation}
\mu_1(N_1/A,N_2/A,T) = \mu_1(N_1,N_2,\sigma(N_1/A,N_2/A,T),T),
\end{equation}
\begin{equation}
\left(\frac{\partial \mu_1}{\partial N_2}\right)_{N_1,A} =\left(\frac{\partial \mu_1}{\partial N_2}\right)_{N_1,\sigma}+ \left(\frac{\partial \mu_1}{\partial \sigma}\right)_{N_1,N_2}\times \left(\frac{\partial \sigma}{\partial N_2}\right)_{N_1,A}.
\label{eq:ConstantSigmaCrossDerivative}
\end{equation}
One recognizes first with the help of (\ref{eq:intermediaire1}), (\ref{eq:intermediaire2}) the identity
\begin{equation}
\left(\frac{\partial \sigma}{\partial N_2}\right)_{N_1,A} = -\frac{\mathcal{A}_2}{A\chi_T}.
\end{equation}
The derivative $(\partial \mu_1/\partial \sigma)$ is directly linked to the specific area $\mathcal{A}_1$ as can be seen, for instance, by examining the differential of the Gibbs free-energy $G = F -\sigma A$,
\begin{equation}
\mathrm{d} G = -A\mathrm{d}\sigma + \mu_1\mathrm{d}N_1 + \mu_2\mathrm{d}N_2\ldots
\end{equation}
and expressing the Maxwell identity 
\begin{equation}
\mathcal{A}_1 = \left(\frac{\partial A}{\partial N_1}\right)_{N_2,\sigma}=-\left(\frac{\partial\mu_1}{\partial \sigma}\right)_{N_1,N_2}.
\end{equation}
Eq.~(\ref{eq:ConstantSigmaCrossDerivative}) is therefore equivalent to 
\begin{equation}
\frac{1}{A}\left(\frac{\partial^2 f}{\partial \rho_1\rho_2}\right) =\left(\frac{\partial \mu_1}{\partial N_2}\right)_{N_1,\sigma}+ \frac{\mathcal{A}_1\mathcal{A}_2}{A\chi_T}.
\label{eq:intermediaire3}
\end{equation}
With some elementary algebra and after simplifications, eq.~(\ref{eq:intermediaire3}) leads to the desired relation
\begin{eqnarray}
\mu_{12} &=& -\frac{k_BT}{\eta A}\nonumber\\
&=& \frac{-k_B T}{A[\rho_1+\rho_2 + \rho_1\rho_2(G_{11}+G_{22}-2G_{12})]}.
\label{eq:expressionForMu12}
\end{eqnarray}
Identification of (\ref{eq:expressionForMu12}) with (\ref{eq:MuCrossDerivative}) finally leads to 
our final expression~(\ref{eq:KBEquationForB}).

\subsection{Contribution of lipid tilt, undulations and inclination modes}

Let us consider the field $\mathbf{M}(\mathbf{r})$ representing the main direction of lipid molecules (director vector) in a leaflet~(Fig~\ref{fig:IllustrationTilt}). We assume first that the membrane is flat and that lipids pivot around a fixed point located on a neutral surface of the leaflet. We also assume that the lipid directors of both leaflets fluctuate independently. Finally, we only illustrate the phenomenon in the pure bilayer situation. The presence of interleaflet tilt correlations is not expected to change the picture. 

In a usual lamellar phase of type $L_{\alpha}$ there is a restoring force that maintains the average value of the parallel projection $\mathbf{M}_{\parallel}$ of $\mathbf{M}$ onto the $x,y$ plane equal to~0. We therefore consider the following expression for the long range elasticity of the tilt: 
\begin{eqnarray}
\mathcal{H} &=& \int_0^{L_x}\dd x\,\int_0^{L_y}\dd y\, \left\lbrace
\frac{K_0}{2}\mathbf{M}_{\parallel}^2
\right\rbrace \nonumber\\
&=& \frac{1}{L_xL_y} \sum_{\mathbf{q}} \frac{K_0}{2}\hat{\mathbf{M}}_{\parallel,\mathbf{q}}\cdot\hat{\mathbf{M}}_{\parallel,-\mathbf{q}} 
\end{eqnarray}
where $\hat{\mathbf{M}}_{\parallel,\mathbf{q}}$ is obtained through eq.~(\ref{eq:FourierTransformCGField}). 
An inhomogeneous director field creates a local fluctuation density equal to 
\begin{equation}
\delta \rho(\mathbf{r}) = -\rho_0\mathrm{div}(\mathcal{D} \mathbf{M}_{\parallel}(\mathbf{r})),
\end{equation}
or equivalently in Fourier space 
\begin{equation}
\hat\rho_{\mathbf{q}} = -\mathcal{D} \mathbf{q}\cdot\hat{\mathbf{M}}_{\parallel,\mathbf{q}},
\end{equation}
leading to the following quadratic fluctuations 
\begin{eqnarray}
    \mean{{\hat\rho}_{\mathbf{q}}\hat{\rho}_{-\mathbf{q}}} &=& \rho_0^2\mathcal{D}^2 \left\lbrace q_x^2 \mean{\hat{\mathbf{M}}_{x,\parallel,\mathbf{q}}\hat{\mathbf{M}}_{x,\parallel,-\mathbf{q}}}\right.\nonumber\\
    & & \;\;\;\left. +q_y^2 \mean{\hat{\mathbf{M}}_{y,\parallel,\mathbf{q}}\hat{\mathbf{M}}_{y,\parallel,-\mathbf{q}}} \right\rbrace\nonumber\\
    &=& \frac{\rho_0^2\mathbf{q}^2 \mathcal{D}^2}{2}\mean{\hat{\mathbf{M}}_{\parallel,\mathrm{q}}\cdot\hat{\mathbf{M}}_{\parallel,-\mathrm{q}}}
\end{eqnarray}
in the isotropic fluctuations case. As $\mathbf{M}$ is a real vector field, its Fourier transform obeys $\hat{\mathbf{M}}_{\parallel,-\mathbf{q}}=\hat{\mathbf{M}}^{*}_{\parallel,\mathbf{q}}$. The coefficients $\mathbf{q}$ and $-\mathbf{q}$ thus contains the same information. One therefore introduces a truly independent subset $\mathcal{Q}$ of vectors $\mathbf{q}$:
\begin{equation}
\mathcal{Q} = \left\lbrace (q_x,q_y), q_x > 0\,\mathrm{or}\, (q_x=0\,\mathrm{and}\, q_y> 0)\right\rbrace
\end{equation}
and the real and imaginary parts of 
$\hat{\mathbf{M}}_{\mathbf{q}}$ such that 
\begin{eqnarray}
\mathcal{H} &=& \frac{K_0}{L_xL_y} \sum_{\mathbf{q}\in \mathcal{Q}} \left\lbrace \mathrm{Re}(\hat{M}_{x,\parallel,\mathbf{q}})^2 +\mathrm{Im}(\hat{M}_{x,\parallel,\mathbf{q}})^2 \right. \nonumber\\
& & \left. + \mathrm{Re}(\hat{M}_{y,\parallel,\mathbf{q}})^2 +\mathrm{Im}(\hat{M}_{y,\parallel,\mathbf{q}})^2 \right\rbrace\nonumber\\
& & + \frac{K_0}{2L_xL_y} \left\lbrace\mathrm{Re}(M_{x,\parallel,\mathbf{0}})^2 + \mathrm{Re}(M_{y,\parallel,\mathbf{0}})^2 \right\rbrace.
\end{eqnarray}

The above expression of the energy gives directly the quadratic fluctuations of the modes 
\begin{eqnarray}
\mean{\mathrm{Re}(\hat{M}_{x,\parallel,\mathbf{q}})^2}
= \mean{\mathrm{Im}(\hat{M}_{x,\parallel,\mathbf{q}})^2} & =& \mean{\mathrm{Re}(\hat{M}_{y,\parallel,\mathbf{q}})^2}\nonumber\\
&=& \mean{\mathrm{Im}(\hat{M}_{y,\parallel,\mathbf{q}})^2}\nonumber\\
&=& \frac{k_B T L_x L_y}{2K_0}
\end{eqnarray}
from which we obtain the desired relation 
\begin{eqnarray}
\mean{\hat{\mathbf{M}}_{\parallel,\mathrm{q}}\cdot\hat{\mathbf{M}}_{\parallel,-\mathrm{q}}} &=& \frac{k_B T}{K_0} L_xL_y\nonumber\\
\mean{{\hat\rho}_{\mathbf{q}}\hat{\rho}_{-\mathbf{q}}} &=& N\rho_0 k_B T \frac{\mathcal{D}^2 \mathbf{q}^2}{2 K_0}\nonumber\\
& =& N S_{\mathrm{tilt}}(q).
\end{eqnarray}
$N$ representing the number of lipid considered in the leaflet or the bilayer.
The conclusion is that the contribution of tilt modes to the structure factor vanishes in the $q\to 0$ limit.

\subsection{Contribution of undulations}

We now consider the contribution of the membrane undulations. The separation between local tilt and membrane inclination is arbitrary as discussed by Watson et al.~\cite{2011_Watson_Brown_2, 2012_Watson_Brown}. We nevertheless adopt the view that there exist a Helfrich manifold $z(\mathbf{r})$ that provides a bilayer normal vector $\mathbf{n}$ with respect to which local tilt is defined. We consider now the consequences of this fluctuating surface.

Treating the membrane as a 2-dimensional manifold with no intra-leaflet tilt amounts to identify the director $\mathbf{M}$ to the bilayer normal vector $\mathbf{n}$. Starting from the elevation function $z(\mathbf{r})$, one deduces the coordinates of the normal vector 
\begin{eqnarray}
\mathbf{n} &=& \frac{1}{\sqrt{1 +z_{,x}^2 + z_{,y}^2}} (-z_{,x}, -z_{,y}, 1)\nonumber\\
& =&  (-z_{,x}, -z_{,y}, 1),
\end{eqnarray}
the second expression being valid in the low surface tilt limit. 

As in the previous section, we assume that leaflets slide freely relative to each other and that lipid molecule pivot around their neutral surface. This again corresponds to a local concentration excess 
\begin{equation}
\delta \rho = -\rho_0 \mathrm{div}(\mathcal{D}\mathbf{n}_{\parallel}) = \rho_0 \mathcal{D} \Delta z(\mathbf{r}).
\end{equation}
The induced concentration fluctuation depends on the Laplacian of the elevation, and is also directly connected to the surface mean curvature. 

One therefore must compute in Fourier space the correlation function 
\begin{equation}
\mean{\hat{\rho}_{\mathbf{q}}\hat{\rho}_{-\mathbf{q}}}
= \rho^2 \mathcal{D}^2 (\mathbf{q}^2)^2 \mean{\hat{z}_{\mathbf{q}} \hat{z}_{-\mathbf{q}}}.
\end{equation}
The calculation is standard. One postulates a Helfrich curvature energy
\begin{equation}
\mathcal{H} = \int_0^{L_x}\dd x\,\int_0^{L_y}\dd y\, \left\lbrace\frac{\kappa}{2}(\Delta z)^2\right\rbrace
\end{equation}
and expresses it in Fourier space thanks to the relation
\begin{equation}
z(\mathbf{r}) = \frac{1}{L_xL_y}\sum_{\mathbf{q}}\hat{z}_{\mathbf{q}}e^{-i\mathbf{q}\cdot\mathbf{r}}
\end{equation}
with the same set of reciprocal vectors $\mathbf{q}$ as in the lipid tilt calculation. To simplify further the calculation, one notices that the mean elevation of the bilayer should be constant and can be set to 0 in full generality, leading to $\hat{z}_{\mathbf{0}}=0$.  
\begin{eqnarray}
\mathcal{H} &=& \frac{1}{(L_xL_y)^2} \int_{\mathcal{S}} \dd \mathbf{r}\, \sum_{\mathbf{q}}\sum_{\mathbf{q}'}
\frac{\kappa}{2} (-q^2)\hat{z}_{\mathbf{q}} (-q'^2)\hat{z}_{\mathbf{q'}} e^{-i\mathbf{q}\cdot\mathbf{r}-i\mathbf{q}'\cdot\mathbf{r}}\nonumber\\
&=& \frac{1}{L_xL_y} \sum_{\mathbf{q}} \left[\frac{\kappa}{2} q^4\right]\hat{z}_{\mathbf{q}}\hat{z}_{-\mathbf{q}} \nonumber\\
&=& \frac{\kappa}{L_xL_y}\sum_{\mathbf{q}\in \mathcal{Q}} q^4 \left(\mathrm{Re}(\hat{z}_{\mathbf{q}})^2+\mathrm{Im}(\hat{z}_{\mathbf{q}})^2\right),
\label{eq:HelfrichReciprocalSpace}
\end{eqnarray}
from which one deduces that 
\begin{equation}
\mean{\mathrm{Re}(\hat{z}_{\mathbf{q}})^2}
= \mean{\mathrm{Im}(\hat{z}_{\mathbf{q}})^2} 
= \frac{k_B T L_x L_y}{2\kappa q^4}
\end{equation}
and finally 
\begin{equation}
\mean{\hat{z}_{\mathbf{q}}\hat{z}_{-\mathbf{q}}} = \frac{k_B T L_x L_y}{\kappa q^4}.
\end{equation}

The undulations contribute to a finite apparent structure factor in the hydrodynamic limit
\begin{eqnarray}
NS_{\perp}(q) &=& \mean{\hat{\rho}_{\mathbf{q}}\hat{\rho}_{-\mathbf{q}}} = \rho_0^2 \mathcal{D}^2 q^4 \mean{\hat{z}_{\mathbf{q}}\hat{z}_{-\mathbf{q}}}\nonumber\\
&=& N \rho_0 \mathcal{D}^2\frac{k_B T}{\kappa}.
\end{eqnarray}
Assuming that longitudinal and undulation contributions are Gaussian and independent lead eventually to the expression~(\ref{eq:AllFluctuations}). In the presence of membrane tension $\sigma$, one would have obtained 
\begin{equation}
\mean{\hat{z}_{\mathbf{q}}\hat{z}_{-\mathbf{q}}} = \frac{k_B T L_x L_y}{\sigma q^2+\kappa q^4}.
\end{equation}
and the corresponding $S_{\perp}(q)$ would vanish in the low $q$ limit.

\subsection{Membrane inclinations}

Any projection of a tilted membrane onto a planar surface tends to increase the apparent density. The ratio between projected and true surface is a geometrical metric factor $\sqrt{1+z_{,x}^2+z_{,y}^2}$. Such effect was considered for instance by Reister and Seifert to discuss the possible slowing down of the apparent diffusion of membrane inclusions~\cite{2005_Reister_Seifert}.

A locally flat but tilted membrane with density $\rho(x,y)=\rho_0+\delta\rho(x,y)$ appears to have a projected density 
\begin{eqnarray}
\rho_{\mathrm{app}}&=&\sqrt{1+z_{,x}^2+z_{,y}^2}\rho(x,y)\nonumber\\
&\simeq& \rho_0 \frac{(\nabla z)^2}{2} + \delta\rho(x,y)+\rho_0.
\end{eqnarray}
The projection creates a purely geometric term $\rho_0 (\nabla z)^2/2$ which gives rise to an apparent structure factor $S_{\mathrm{proj}}(q)$:
\begin{equation}
N S_{\mathrm{proj}}(q) = \mean{\hat\rho_{g,\mathbf{q}}\hat\rho_{g,-\mathbf{q}}}
\end{equation}
with $\hat\rho_{g,-\mathbf{q}}$ the Fourier transform of $\rho_0 (\nabla z)^2/2$.

\begin{equation}
\frac{(\nabla z)^2}{2} = \frac{1}{2L_x^2L_y^2} \sum_{\mathbf{q}_1}\sum_{\mathbf{q}_2} (-i\mathbf{q}_1)\cdot(-i \mathbf{q}_2) e^{-i(\mathbf{q}_{1}+\mathbf{q}_{2})\cdot\mathbf{r}} \hat{z}_{\mathbf{q}_1}\hat{z}_{\mathbf{q}_2};
\end{equation}

\begin{equation}
\hat\rho_{g,\mathbf{q}} = \frac{\rho_0}{2L_xL_y}\sum_{\mathbf{q}_1} \hat{z}_{\mathbf{q}_1}\hat{z}_{\mathbf{q}-\mathbf{q}_1} \big(\mathbf{q}_1\cdot(\mathbf{q}_1-\mathbf{q})\Big).
\end{equation}
One checks first that $\mean{\hat\rho_{g,\mathbf{q}}}=0$ for a thermalized Helfrich Hamiltonian $\mathbf{q}\neq 0$. By construction $\hat{z}_{\mathbf{q}=0}=0$, and then 
\begin{equation}
\mean{\hat{z}_{\mathbf{q}_1}\hat{z}_{\mathbf{q}-\mathbf{q}_1}} = \frac{k_B  T L_xL_y}{\kappa q^4} \delta_{\mathbf{q}_1,\mathbf{q}_1-\mathbf{q}} =0.
\end{equation}
The $\mathbf{q}=0$ term 
\begin{equation}
\int_{\mathcal{S}} \dd\mathbf{r}\, \frac{(\mathbf{\nabla} z)^2}{2}
= \frac{k_B T}{2\kappa} \sum_{\mathbf{q}\neq 0} \frac{1}{q^2}
\end{equation}
is a diverging series giving rise to a logarithmic term $k_B T\ln(L_xL_y)/(8\pi\kappa)$. it corresponds to the celebrated ratio between apparent and projected area of a fluctuating membrane~\cite{1973_Helfrich_2, 1981_Kwok_Evans, 2005_denOtter}. 

The calculation of $\mean{\hat\rho_{g,\mathbf{q}}\hat\rho_{g,-\mathbf{q}}}$ can be done with usual techniques for Gaussian fluctuating fields. We find 
\begin{eqnarray}
\mean{\hat\rho_{g,\mathbf{q}}\hat\rho_{g,-\mathbf{q}}}&=&
\frac{\rho_0^2(k_B T)^2}{2\kappa^2} \sum_{\mathbf{q}_1} \frac{(\mathbf{q}_1\cdot(\mathbf{q}-\mathbf{q}_1))^2}{q_1^4(q-q_1)^4}\\
&=&\frac{N^2(k_B T)^2}{32\pi^4 \kappa^2}\times\nonumber\\
& & \sum_{\stackrel{n'_x\neq 0;n'_y\neq 0}{n'_x\neq n_x; n'_y\neq n_y}} \frac{[n'_x(n_x-n'_x) + n'_y(n_y-n'_y)]^2}{[{n'_x}^2+{n'_y}^2]^2[(n_x-{n'_x})^2+(n_y-{n'_y})^2]^2}.\nonumber
\end{eqnarray}
A quick numerical estimate of the double series led us to the conclusion that it should be numerically bounded by a constant close to $C\simeq 6$. Thus the contribution to $S_{\mathrm{proj}}(q)$ is of order 
\begin{equation}
S_{\mathrm{proj}}(q\to 0) < \frac{C N}{32\pi^4} \left(\frac{k_BT}{\kappa}\right)^2 \simeq 0.005.
\end{equation}
for $N\sim 1000$ and $\kappa = 20 k_BT$, typical values for our system. We are lucky that lipid bilayers are quite rigid, leading to a very small value of $k_B T/\kappa$. In addition, the geometrical projection term does not affect in principle the relative mixing of the components and should not contribute to $B$. In the current study we ignore the contribution of this term, but notice that the linear dependence in $N$ could make it relevant for larger or softer systems. 

\providecommand{\noopsort}[1]{}\providecommand{\singleletter}[1]{#1}%

\end{document}